\documentclass[fleqn,usenatbib,useAMS]{mnras}
\usepackage[T1]{fontenc}
\usepackage{ae,aecompl}
\usepackage{graphicx}
\usepackage{verbatim}
\usepackage{amssymb}
\usepackage{ulem} %to strike a line
\usepackage{color} %color text
\usepackage{url}
\usepackage{upgreek}
\usepackage{txfonts} % an approximation of MNRAS font that is accepted by arXiv.org

\makeindex \citeindextrue

\newcommand{\Alfven}{ Alfv\'{e}n }
\newcommand{\Lf}{{Lorentz factor}}
\newcommand{\Bf}{{magnetic field}}
\newcommand{\Bfs}{{magnetic fields}}
\newcommand{\BH}{{black hole}}

\title[Polarization swings  in blazars]{Polarization swings  in blazars}
\author[M. Lyutikov \& E. Kravchenko]{Maxim Lyutikov$^{1}$\thanks{Contact e-mail: \href{mailto:lyutikov@purdue.edu}{lyutikov@purdue.edu}} and Evgeniya V. Kravchenko$^{2}$\\
$^{1}$Department of Physics, Purdue University, 525 Northwestern Avenue, West Lafayette, IN 47907-2036, USA\\
$^{2}$Lebedev Physical Institute, Astro Space Center, Profsoyuznaya 84/32, Moscow 117997, Russia\\}

\date{Accepted ... Received ...; in original form ...}
\pubyear{2017}

\begin{document}
\label{firstpage}
\pagerange{\pageref{firstpage}--\pageref{lastpage}} 
\maketitle

\begin{abstract}
We {present} a model of blazar variability that can both reproduce  smooth large polarization angle swings, and at the same time allow for the seemingly random  behavior of synchrotron fluxes, polarization fraction  and, occasionally,  $\uppi/2$  polarization jumps.  
We associate blazar flaring activity with a jet carrying helical \Bfs\ and propagating  along a variable direction (and possibly with a changing bulk \Lf). 
The model predicts that for various jet trajectories  (i) EVPA can experience large smooth {temporal} variations while at the same time  polarization fraction ($\Pi$) can be highly variable; (ii)  $\Pi \sim 0$ near sudden EVPA jumps of  90$^{\circ}$, but can also remain constant for large, smoother EVPA swings; (iii)  the total angle of EVPA rotation can be arbitrary large; 
(iv)  intensity $I$  is usually maximal at points of fastest EVPA changes, but can have a minimum.  Thus, even for a regular, deterministic motion of a steadily emitting jet  the  observed properties  can vary in a non-monotonic  and/or seemingly stochastic way.  Intrinsic fluctuations of the emissivity will further complicated the intensity profiles, but are expected to preserve the polarization structure.
\end{abstract}

\begin{keywords}
{polarization --- galaxies: active --- galaxies: magnetic fields --- galaxies: jets}
\end{keywords}

\section{Correlated flux and polarization variations -- observational overview and theoretical models}

Blazars -- a sub-class of active galactic nuclei (AGN) -- have the orientation of their jets close to the line of sight (LoS). This makes their non-thermal radiation to be highly relativistically beamed. 
Their linear fractional polarization reaches values up to 50 per cent \citep[e.g.][]{2005AJ....130.1389L} suggesting  the presence of highly ordered  magnetic fields in their compact regions \citep{2005MNRAS.360..869L}.
Furthermore, observed behavior of polarization degree and angle suggest helical shape of these magnetic fields \citep[e.g.][]{1999NewAR..43..691G,2005MNRAS.356..859P,2005MNRAS.360..869L}.

Blazars are observed to show high variability across the electromagnetic spectrum \citep[e.g.][]{1989Natur.337..442Q,1997ARAA..35..445U}. 
The evolution of the $\gamma$-ray, optical, radio and polarized fluxes often exhibit seemingly random behavior \citep[e.g.][and references therein]{2013ApJ...768...40L} and in some cases might be represented by a number of isolated, individual events superimposed on a steady processes \citep[e.g.][]{2008Natur.452..966M,2016AA...590A..10K}.
In contrast, the optical electric vector position angle (EVPA) variations often show smooth swings of the linearly polarized radiation, with total rotations up to few radians (see Section \ref{Comparison}  for corresponding examples). 
The noticeable nature of such events attracted special interest, giving birth to a large polarimetric programs, like RoboPol \citep{2014MNRAS.442.1706K,2014MNRAS.442.1693P}, MAPCAT\footnote{\url{www.iaa.es/~iagudo/research/MAPCAT/MAPCAT.html}}, monitoring at Steward observatory \citep{2016Galax...4...27S}, with Kanata optical telescope \citep{2011PASJ...63..639I}, and others.

%BL Lac and example of a case when Pi drops to  zero
%\begin{comment}
\begin{figure}
 \centering
 \includegraphics[width=.93\columnwidth]{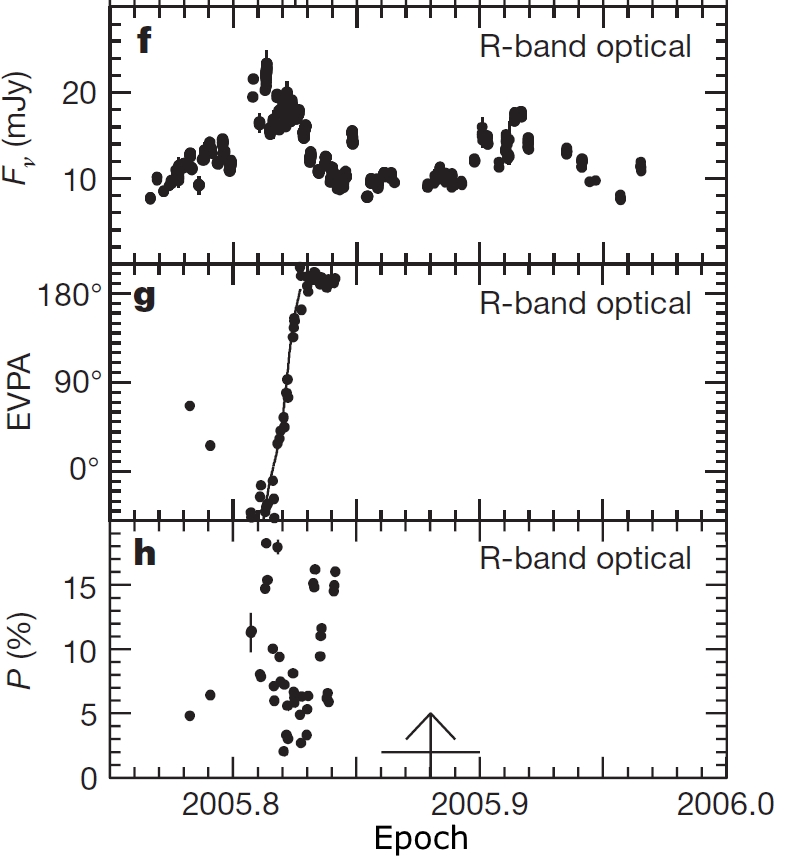}
 \caption{Optical R-band observations of BL~Lac as functions of time: (f) flux density, (g) degree of polarization and (h) EVPA. Figure 2 of \citet{2008Natur.452..966M}.}
 \label{fig_bllac_mars}
\end{figure}
%\end{comment}

Apparent similarities of optical flux, degree of polarization and EVPA during these events, detected in  different sources and their different flaring states, suggests a common mechanism being responsible for such a behavior.
The general pattern includes high variability of polarization $\Pi$,  {which can be both higher and lower during the middle of EVPA swing, but} then recovers back to the initial value, smooth and continuous change of polarization angle and peaked behavior of optical flux density.
Figure~\ref{fig_bllac_mars} presents sample of such behavior.
%The highest amplitude gamma-ray flares are connected with the large EVPA swings, as well as optical and sometimes radio flaring activities.
%Blazars are known to be emitters of $\gamma$-rays~\citep{2015ApJS..218...23A}.
All EVPA rotation events (hereafter we refer to optical EVPA) have been detected to date only in the $\gamma$-ray loud objects, suggesting physical relation of optical and $\gamma$-ray emission sites. 
%but not all of them accompanied by $\gamma$-ray flares \citep[e.g.,][]{2015MNRAS.453.1669B}.
\citet{2015MNRAS.453.1669B} note that these EVPA rotations may be produced via both random walk processes \citep[e.g.][]{2014ApJ...780...87M} and deterministic processes,
%Meanwhile observed correlation of the optical and $\gamma$-ray emission \citep[][]{2010Natur.463..919A,rani_etal13} provides evidence of cospatiality of the regions, where EVPA swings occur.
% \citep[][]{2010Natur.463..919A,rani_etal13} and sometimes radio flaring activities \citep{2008Natur.452..966M,2010Natur.463..919A,2015MNRAS.453.1669B}.
while the latter can be connected with the strongest $\gamma$-ray flares \citep[e.g.][]{2010Natur.463..919A,2014AJ....148...42M,2015MNRAS.453.1669B,2016AA...590A..10K}.
%Though observations show that only strongest $\gamma$-ray flares are accompanied by EVPA swings \citep[e.g.,][]{2010Natur.463..919A,2014AJ....148...42M,2015MNRAS.453.1669B,2016AA...590A..10K}.

A number of models attempt to explain the polarization behavior \citep[e.g.][]{1982ApJ...260..855B,1985ApJ...289..188K,2014ApJ...780...87M,2014MNRAS.437.3405L,2014ApJ...789...66Z}. The key approaches include  many-zone emission models  (in an attempt to explain nearly random behavior of some jet properties) and models that rely on regularly evolving jet parameters (in an attempt to explain smooth variations of other parameters, like the EVPA). For example, \citet{2016AA...590A..10K} concluded that during the flare state a deterministic process governs the polarization variation, while at low-brightness state polarization is more random. The  models  of \citet{2010IJMPD..19..701N,1982ApJ...260..855B} are, perhaps, the closest to the present model.

\section{The model: jet with helical magnetic field propagating along a variable direction}

\begin{figure}
\includegraphics[width=1.4\columnwidth]{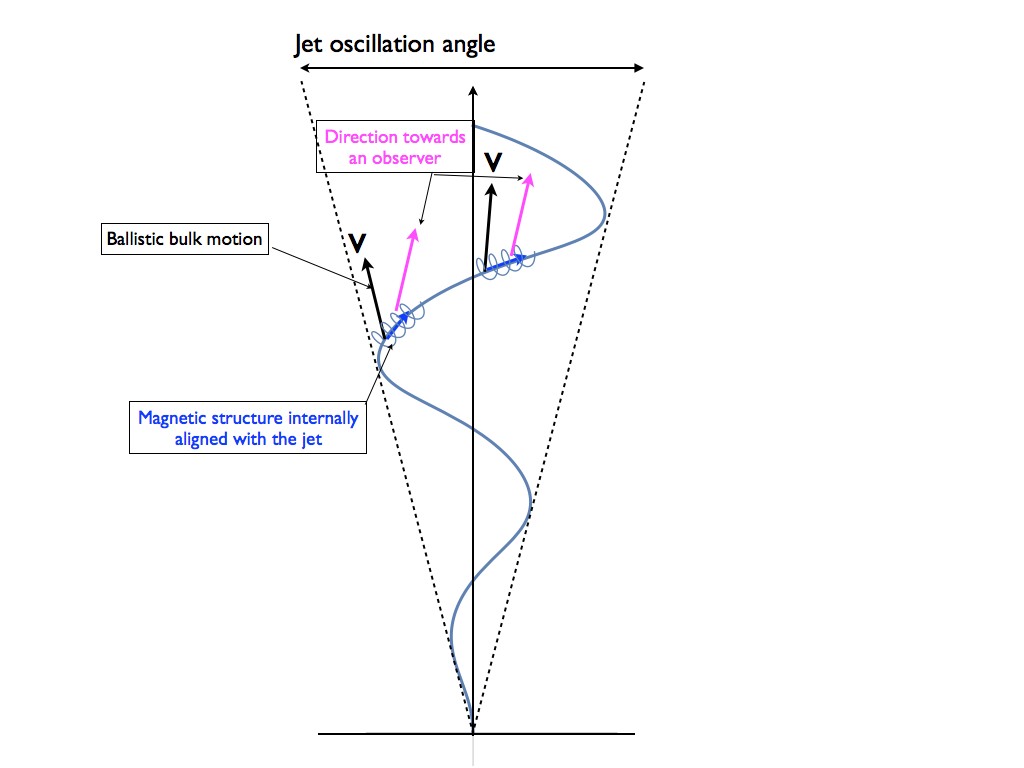}
\caption{Schematic representation of the model. The jet is emitted along a variable direction (defined, e.g.\ by the opening angle of the planar motion, jets' oscillation angle). The internal  helical structure of the \Bf\ within the jet is aligned with the local jet direction and changes with time.} 
\label{Picture-Main}
\end{figure}

In this paper we present a model, showed in Fig. \ref{Picture-Main} - a jet propagating along a smoothly variable direction carrying helical \Bf\  - which is able to reproduce large smooth variations of the EVPA, yet allow for occasional sudden jumps in EVPA. In addition - and most importantly -  the intensity and  polarization fraction, though produced by a highly deterministic process, show large non-monotonic variations that can be mistaken for a random process. Thus, a highly deterministic set-up of  the model  produces both smooth variation of EVPA and yet allows for some properties of the emission to vary in a non-monotonic way, which can be interpreted as stochastic variation.

%\begin{comment}
%\end{comment}

We model the emitting element as a jet carrying helical  \Bf\, with internal pitch angle $\psi$, propagating with \Lf\  $\gamma_j$. The jet produces polarized synchrotron emission.  
We concentrate on the optically thin region, sufficiently far downstream of the  {place, where jet originates}. In terms of physical location the model is applicable to on sub-parsec to parsec scale regions of the jet.  In the present paper we do not make a separation between the different parts of the spectrum, e.g.\ optical and radio, but outline the general properties of polarized synchrotron emission expected from a jet with variable direction.

%\begin{comment}
\begin{figure*}
 \includegraphics[width=1.45\columnwidth]{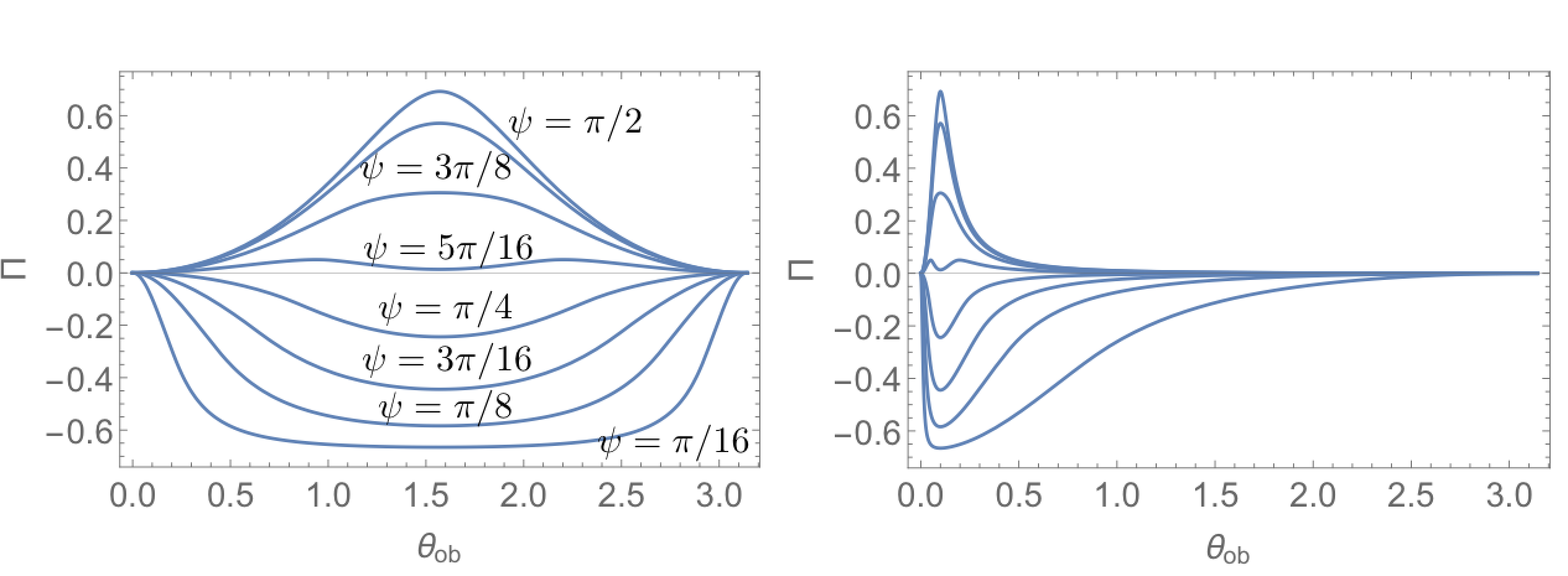}
 \caption{Polarization fraction $\Pi$ for a jet carrying helical \Bf\ as function of viewing angle in comoving (left panel) and observer frames (right panel, $\gamma=10$) for different pitch angles
\citep{2005MNRAS.360..869L}.
Positive values correspond to average polarization along the jet, while negative correspond to polarization perpendicular to the jet. Pitch angels are $0, \uppi/16/\uppi/8 ... \uppi/2$. These values of the pitch angles are used in all the plots below.} 
\label{PiRest}
\end{figure*}
%\end{comment}

Calculations of polarization produced by relativistically moving sources is somewhat complex \citep{1979ApJ...232...34B,2003ApJ...597..998L,2005MNRAS.360..869L}. 
Conventionally (and erroneously for a relativistically moving plasma!), the direction of the observed polarization for optically thin regions and the associated magnetic fields are assumed to be in one-to-one correspondence, being orthogonal to each other, so that some observers choose to plot the direction of the electric vector of the wave, while others plot vectors orthogonal to the electric vectors and call them the direction of the magnetic field. 
\textit{This  is  correct only for non-relativistically  moving optically-thin sources, and thus cannot be applied to AGN jets.}
Since the emission is boosted by the relativistic motion of the jet material, \textit{the  EVPA rotates parallel to the plane containing the line of sight and the plasma velocity vector}, so that \textit{the observed electric field of the wave is not, in general, orthogonal to the observed magnetic field},  \citep{2003ApJ...597..998L,2005MNRAS.360..869L}. 
%For consistency, the results are reproduced in Appendix \ref{PPP}.

We  consider the synchrotron  emission of  an unresolved, thin, circular  cylindrical shell populated by relativistic electrons with a power law distribution and moving uniformly in the axial direction with constant velocity.  
The properties of the synchrotron emission are then determined by \textit{three parameters}: the internal pitch angle of the magnetic field $\psi'$, Lorentz factor of the shell in the laboratory frame  $\gamma_j$ and the viewing angle, $\theta$, which the line of sight to the observer makes with the jet axis in the observer reference frame.
Thus, even for fixed internal parameters of the jet, the resulting polarization signature strongly depends both  on the viewing angle and the jet \Lf\ \citep{2005MNRAS.360..869L}, Fig. \ref{PiRest}.

As a novel feature, we  allow parameters of the model (the viewing angle and the Lorentz factor) to vary smoothly  with time and we  analyze the resulting correlations.
%We  investigate a model wherein  $\gamma$-ray flares are produced by \Bf-carrying jets with temporarily changing directions and/or {\Lf}s. 
As a result of relativistic boosting, at different moments in time the jet is  seen from highly variable directions in the jet's frame. 
The observed intensity, polarization  and EVPA  then experience large variations. Even though these variations are highly correlated, the observed properties show large, seemingly random changes.
We  consider several types of jet variations:  planar oscillating motion, circular motion, jet acceleration and combinations thereof. 

The polarization direction from an unresolved jet can be either along the projection of the jet onto the plane of the sky, or perpendicular to it. Thus, as a jet's direction changes with time, the direction of polarization will also change. Most of the time, the EVPA will either be always along or across the jet. In addition, for a fairly narrow range of internal pitch angles and lines of sight a given jet can show 90$^{\circ}$ EVPA flips. %\sout{, see Fig. \ref{2piover7}.}

Importantly, in this paper we do not address the physical origin of the emission features. Qualitatively, we image that  the jet motion is ballistic, but along time-dependent trajectories/velocities, determined by the changing condition at the location of the jet acceleration, like jet from a firehose. Emission is then produced by a feature moving along the jet. These emission features, propagating along changing direction, then can be modeled  as a jet with variable direction. 
%The changing direction or \Lf\ can be due, e.g.\ to the instabilities developing in the jet  \citep[eg][]{2016MNRAS.tmpL..44T}.

\subsection{Planar motion of the jet direction}
\label{Planar}

\begin{figure}
\centering 
 \includegraphics[width=1.0\columnwidth]{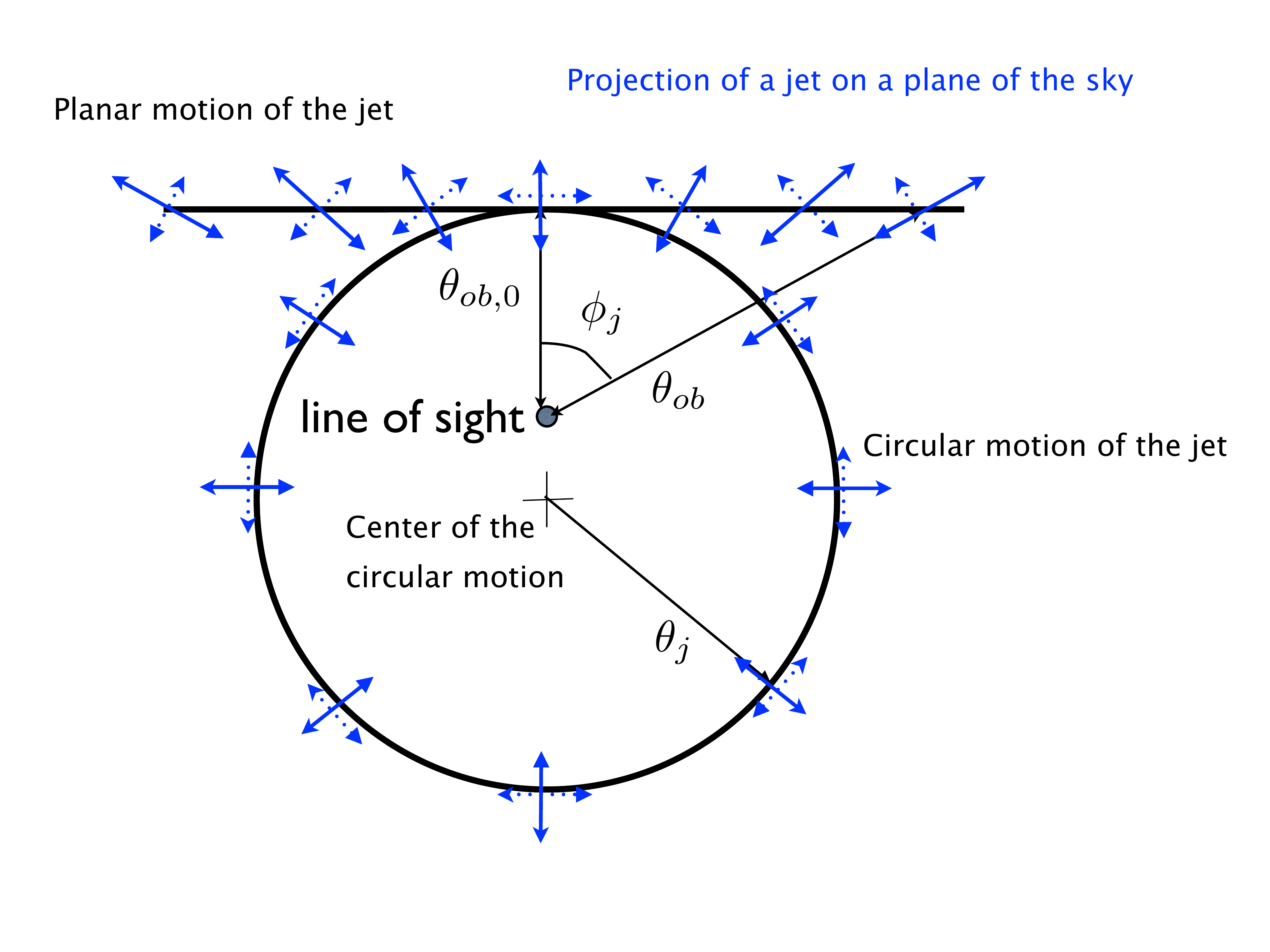}
\caption{Geometry of the model in the plane of the sky. Direction of the jet changes with time executing planar or circular motion. Solid blue arrows represent the projection of the jet on the plane of the sky. At each  point EVPA is either along or perpendicular (dotted arrows) to the projection of a jet on the plane of the sky.  Fastest rate of EVPA change occurs near the closest approach between the line of sight and the jet direction. Depending on the parameters EVPA can flip by 90$^{\circ}$. During such flips polarization will pass through zero. }
\label{geometry-Pi}
\end{figure}

Let us consider how apparent brightness, polarization fraction and EVPA change with time if  the jet's direction executes a regular motion. First, consider planar motion of a jet, so that a jet oscillates with amplitude $\phi_{j, max}= \pm \uppi/2$ making the minimum angle with the LoS $\theta_{ob, 0}$, Fig. \ref{geometry-Pi}.

%\begin{comment}
%\end{comment}

The angle $ \theta_{ob} $ between the LoS  and the jet direction is $\cos  \theta_{ob} =\cos \theta_{ob,0} \cos \phi_{j}$. The 
angle  between a fixed direction and the projection of the jet on the plane of the sky is
\begin{equation}
\sin \phi_{PA} = { \sin \phi_j \over \sqrt{1 - \cos \theta_{ob,0}^2 \cos ^2 \phi_{j}}}
\label{PSJ}
\end{equation}
(the  EVPA can be different from (\ref{PSJ}) by $\uppi/2$). 
 %\bf{Give explanation, which pitch angles correspond different curves.})
The rate of change of $\phi_{PA}$ is:
\begin{equation}
\dot{\phi}_{PA} = \frac{\sin \left(\theta _{{ob},0}\right)}{1-\cos ^2\left(\phi _j\right) \cos ^2\left(\theta _{{ob},0}\right)} \dot{\phi}_j
\label{phiPA}
\end{equation}

%\begin{comment}
\begin{figure*}
\centering 
$\theta_{ob,0}=1/(5 \gamma), \hskip .15 \columnwidth  \theta_{ob,0}=1/\gamma,  \hskip .15 \columnwidth  \theta_{ob,0}=\, 2/\gamma$
\\
 \includegraphics[width=.6\columnwidth]{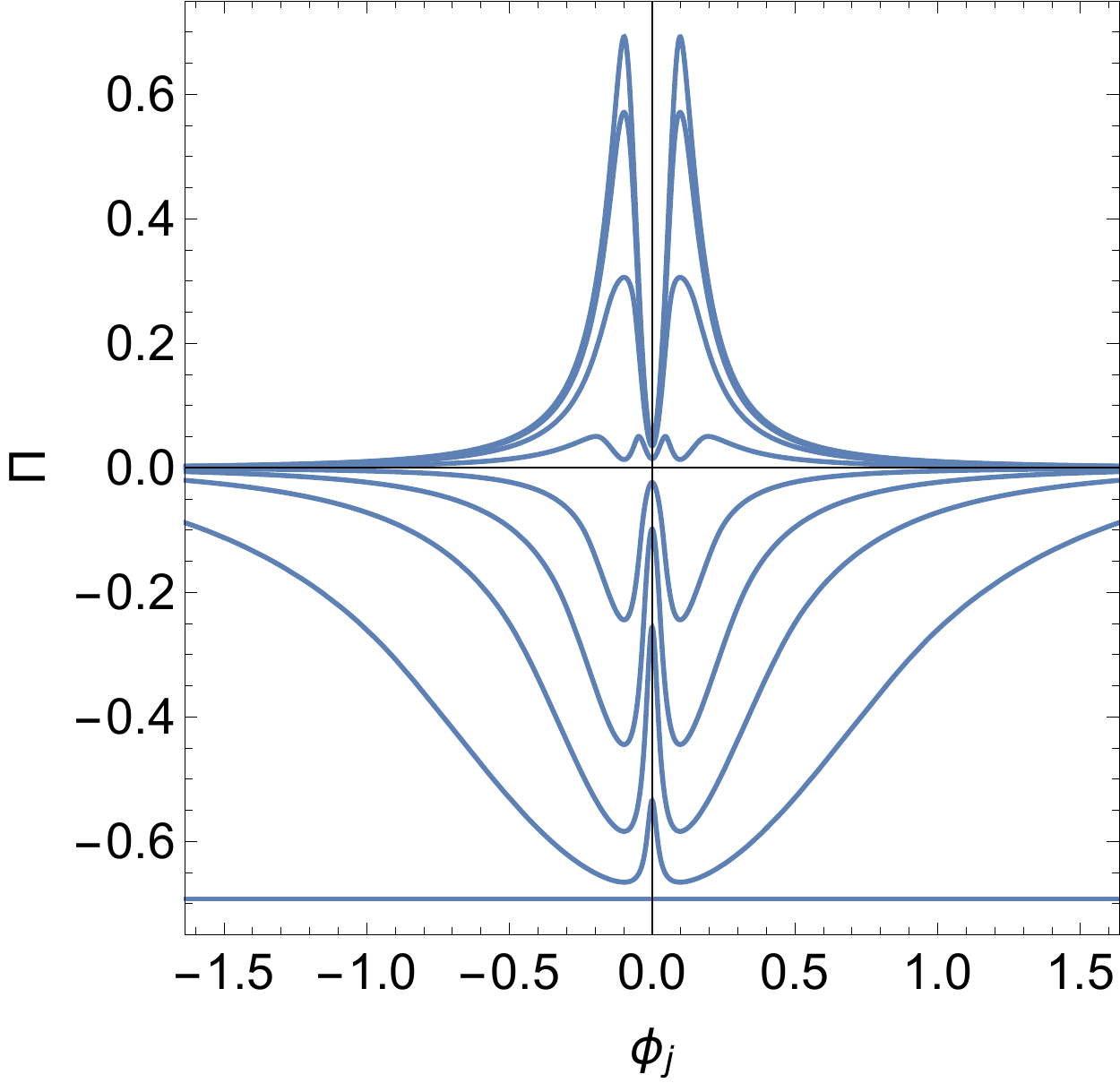}
 \includegraphics[width=.6\columnwidth]{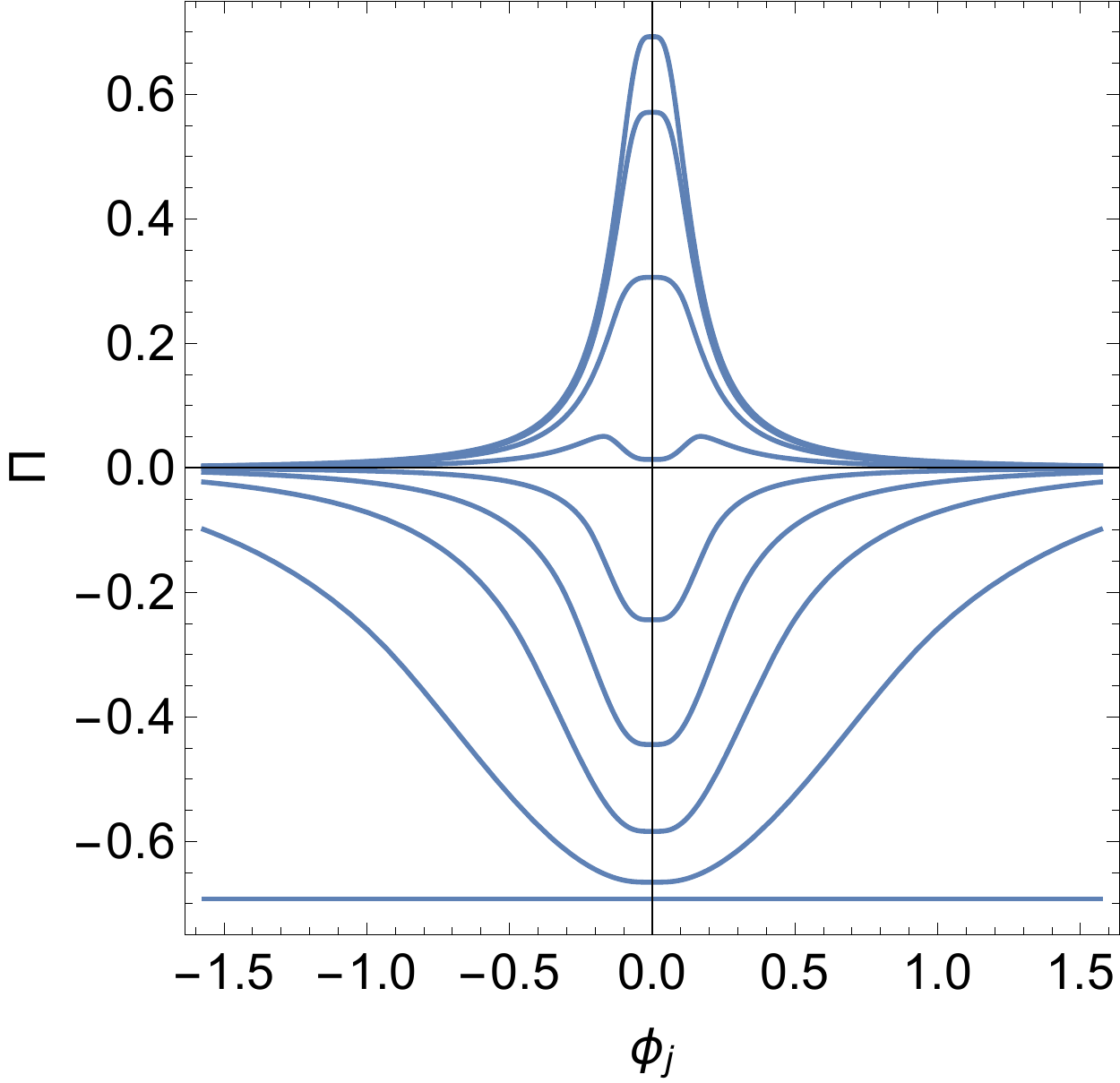}
 \includegraphics[width=.6\columnwidth]{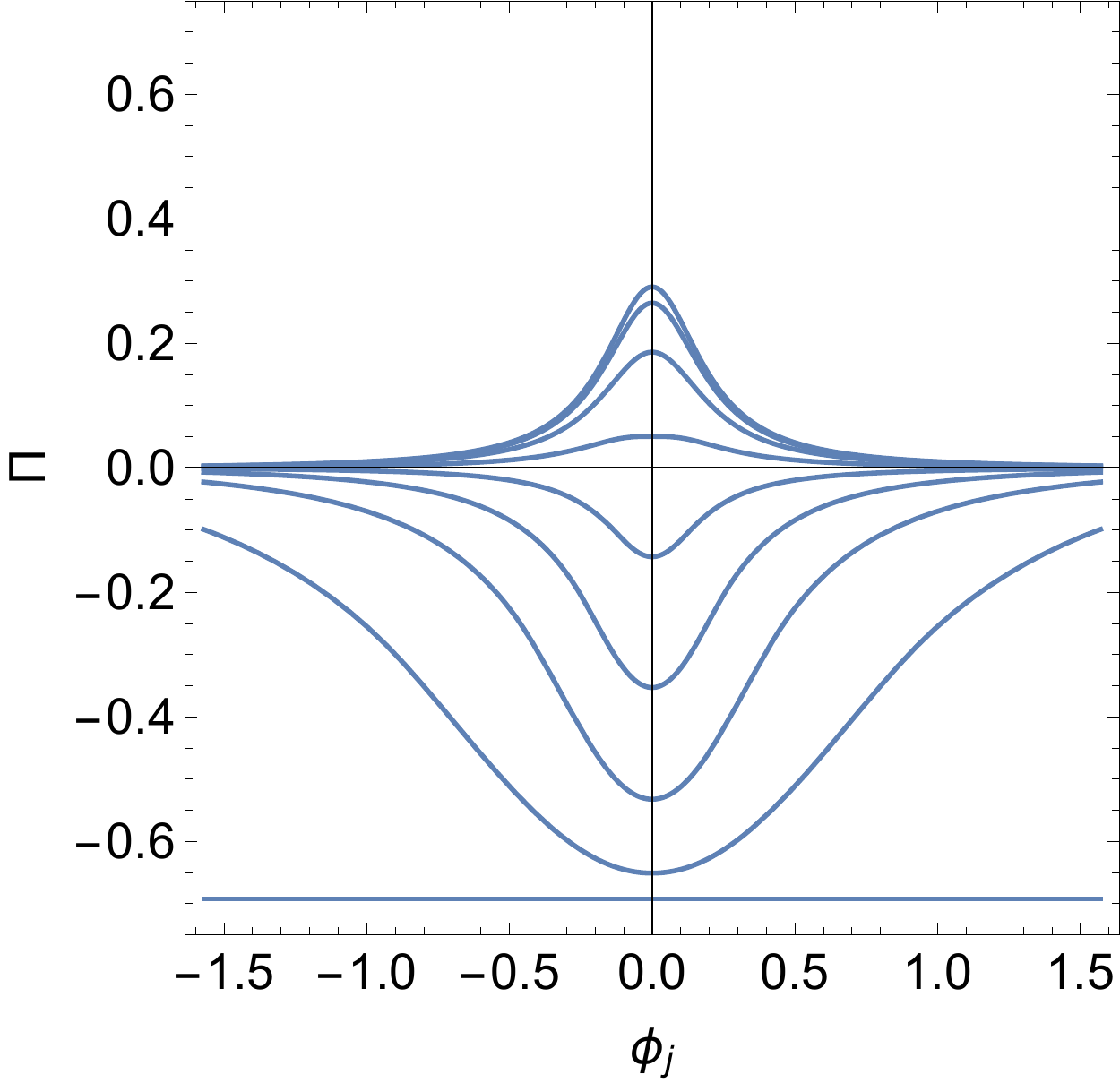}
\\
 \includegraphics[width=.6\columnwidth]{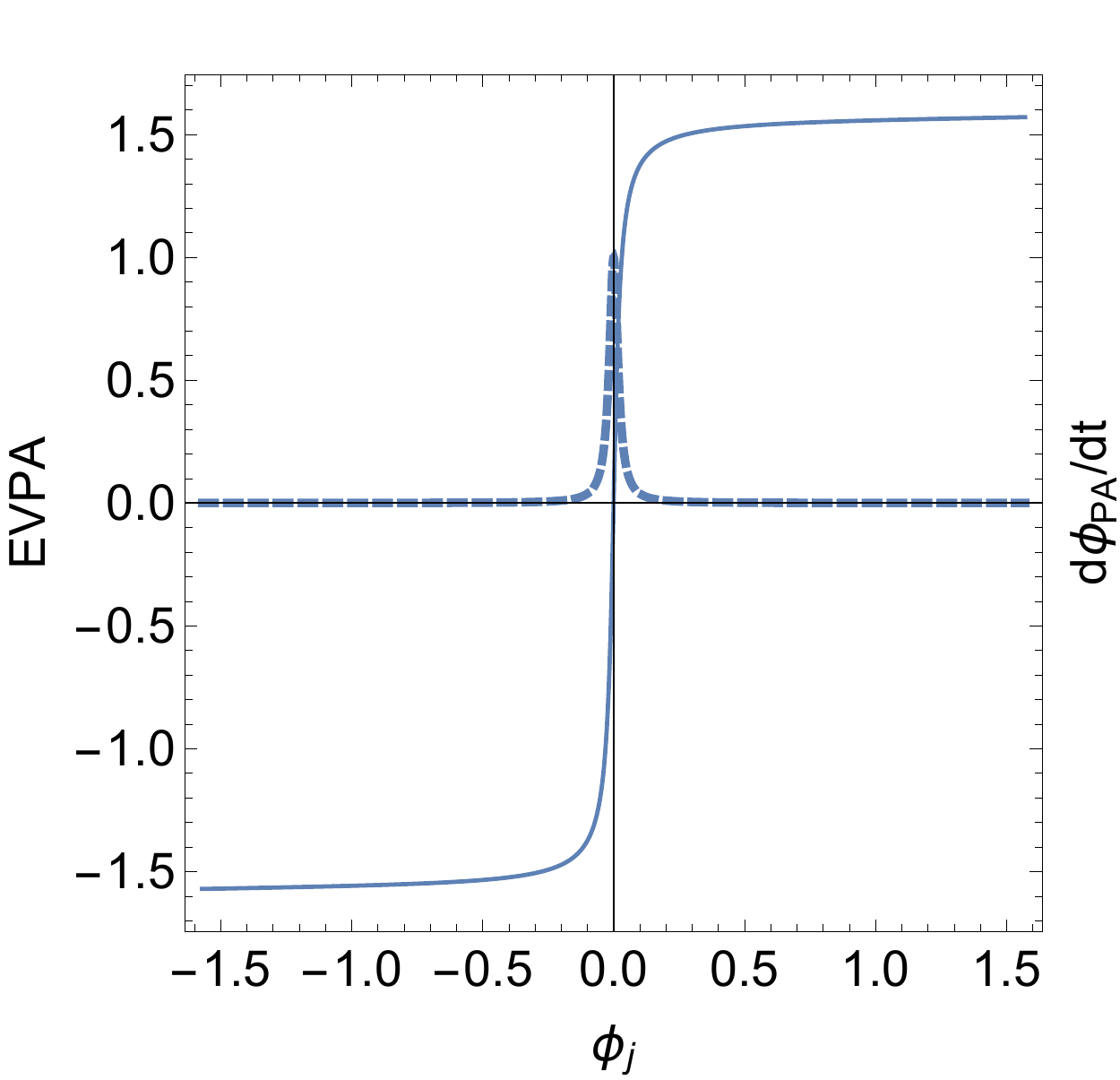}
 \includegraphics[width=.6\columnwidth]{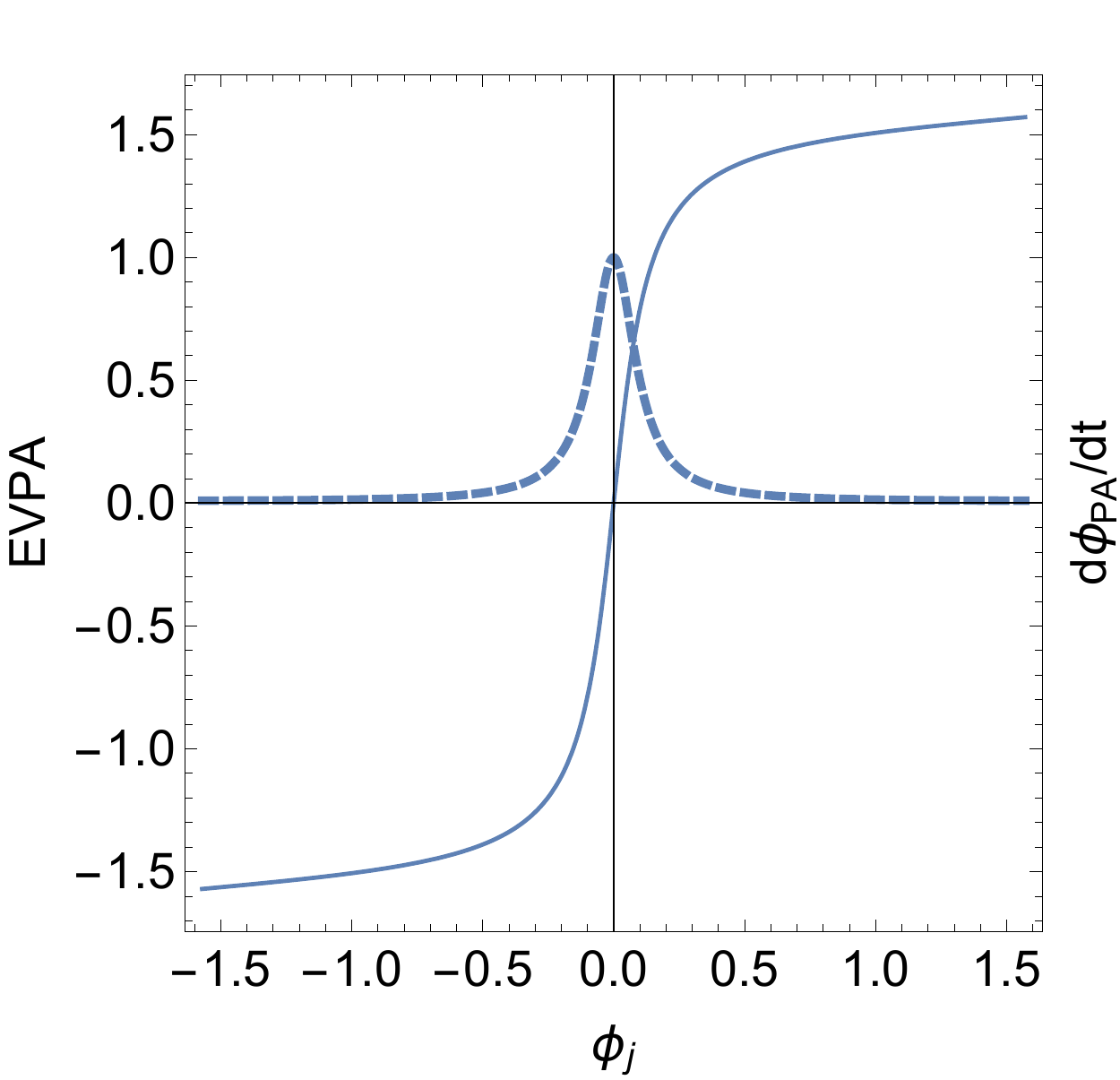}
 \includegraphics[width=.6\columnwidth]{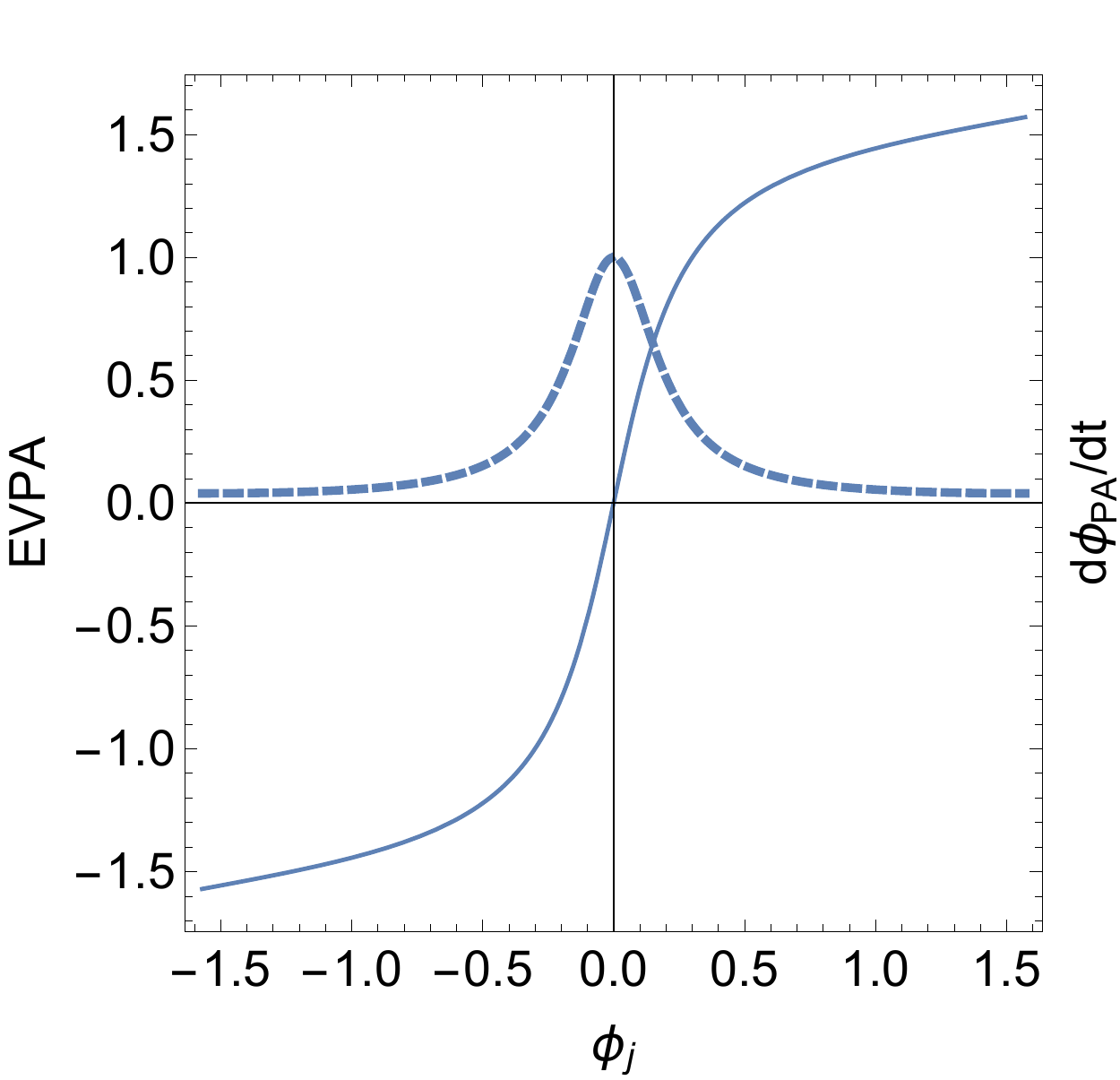}
\caption{Polarization $\Pi$ and EVPA  for a jet executing planar motion. The jet is moving with bulk \Lf\  $\gamma=10$ and is viewed at the minimal viewing angles of  $\theta_{ob,0}=1/(5 \gamma), \, 1/\gamma, \, 2/\gamma$ (left to right columns). \textit{Top row}: $\Pi$ as function of the oscillation angle  for different  intrinsic pitch angles.  \textit{Bottom row}:  EVPA as function of the oscillation angle (solid line). (Here a larger range of angles $\phi_j$ is plotted to show the full periodic behavior of EVPA).  Dashed line: the rate of change of EVPA, $\dot{\phi}_{PA}$ (defined here as the projection of the jet on the plane of the sky - EVPA may differ by $90^\circ$), normalized to the maximal value.} 
\label{Piofphij}
\end{figure*}
%\end{comment}

%\begin{comment}
\begin{figure*}
\centering 
$\theta_{ob,0}=1/(5 \gamma), \hskip .15 \columnwidth  \theta_{ob,0}=1/\gamma,  \hskip .15 \columnwidth  \theta_{ob,0}=\, 2/\gamma$\\
 \includegraphics[width=.6\columnwidth]{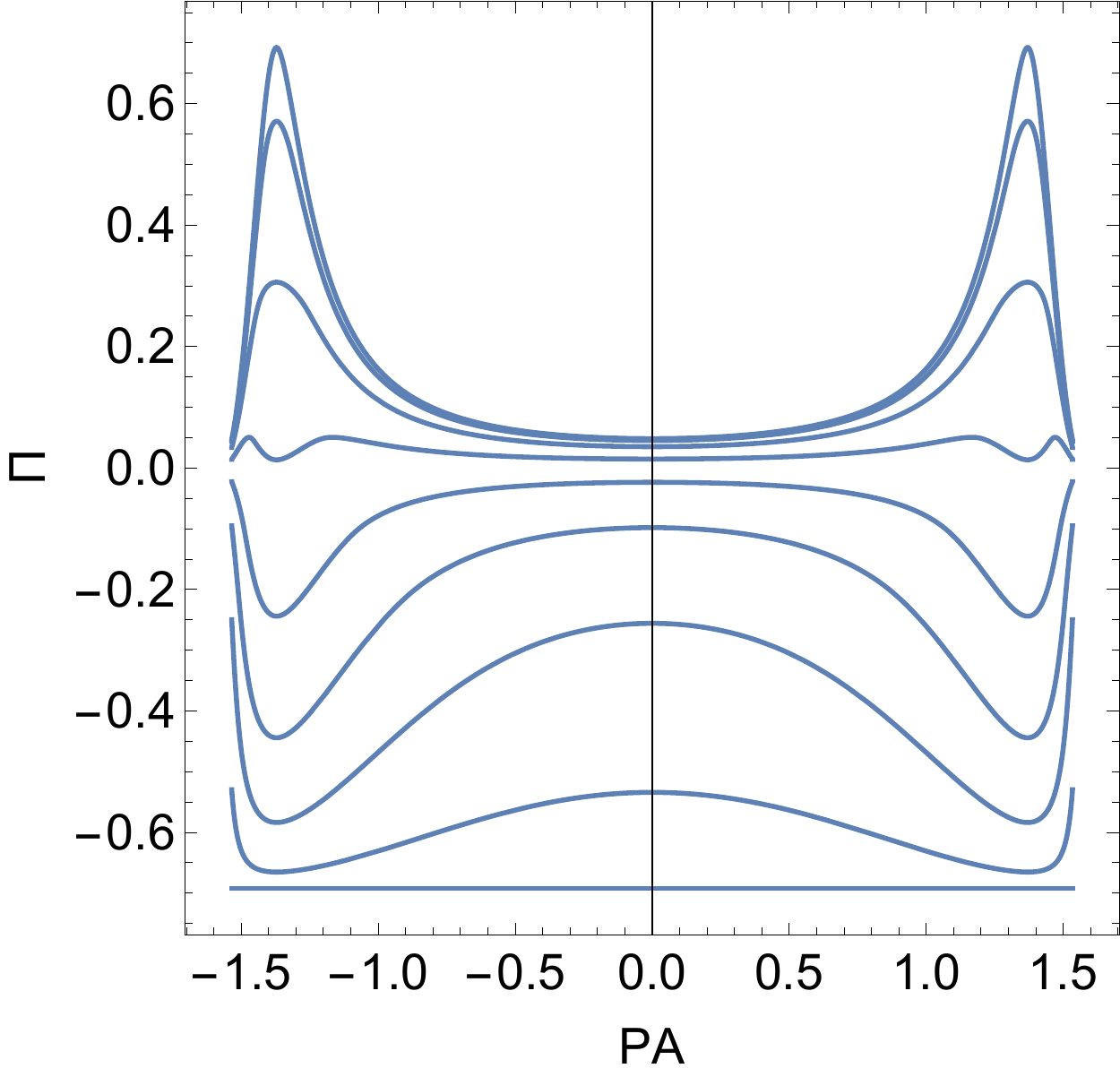}
 \includegraphics[width=.6\columnwidth]{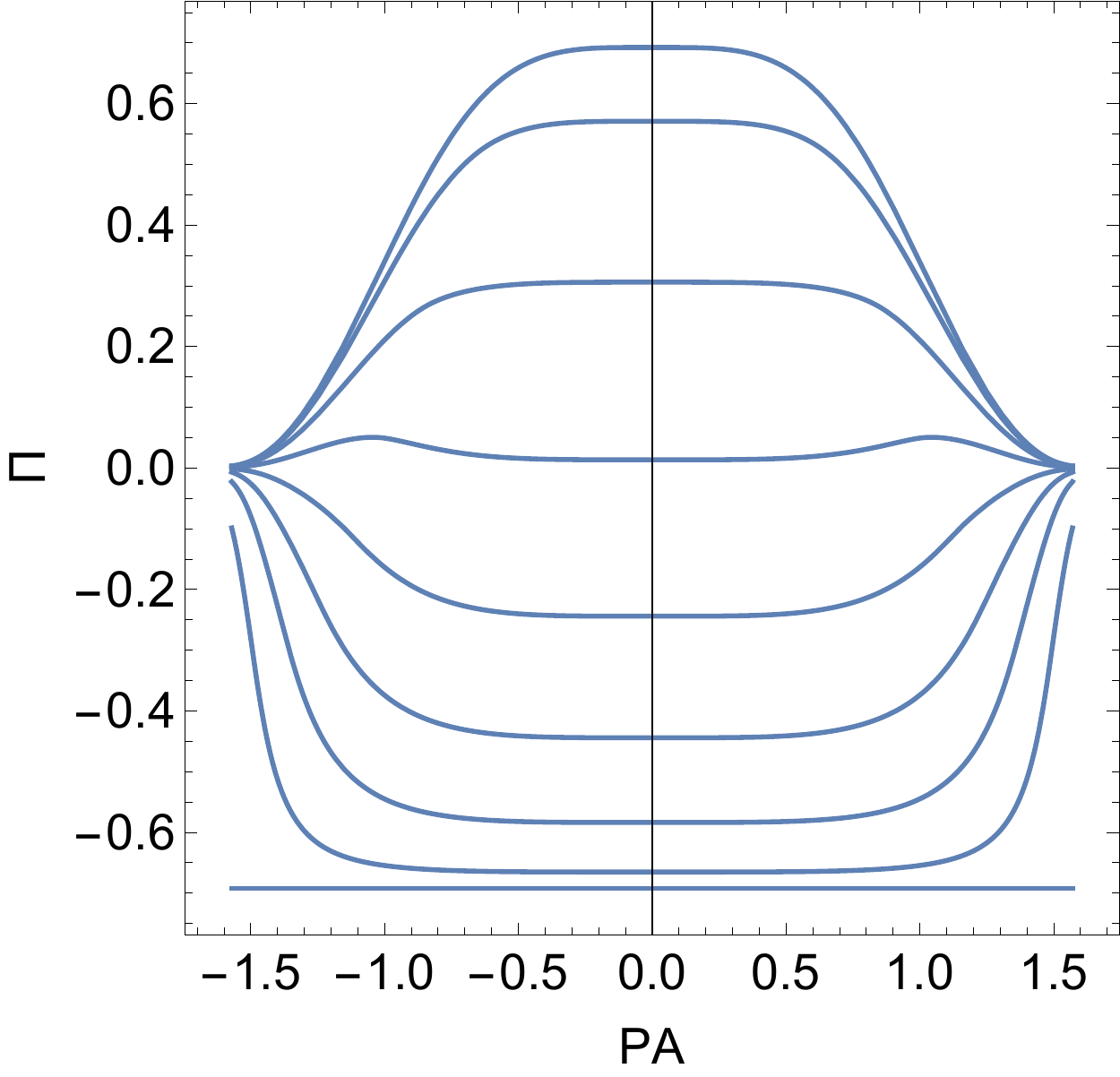}
 \includegraphics[width=.6\columnwidth]{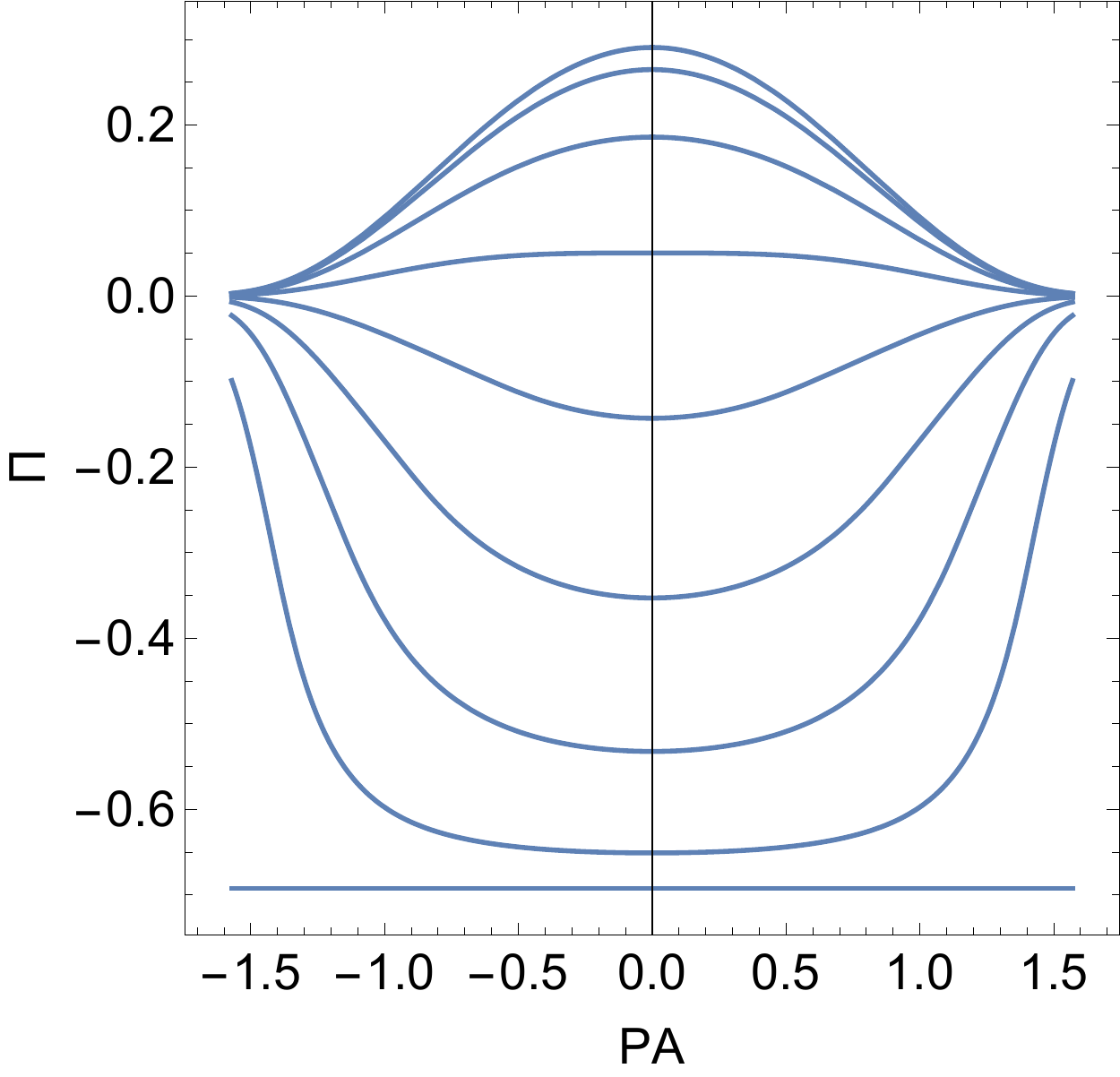}\\
 \includegraphics[width=.6\columnwidth]{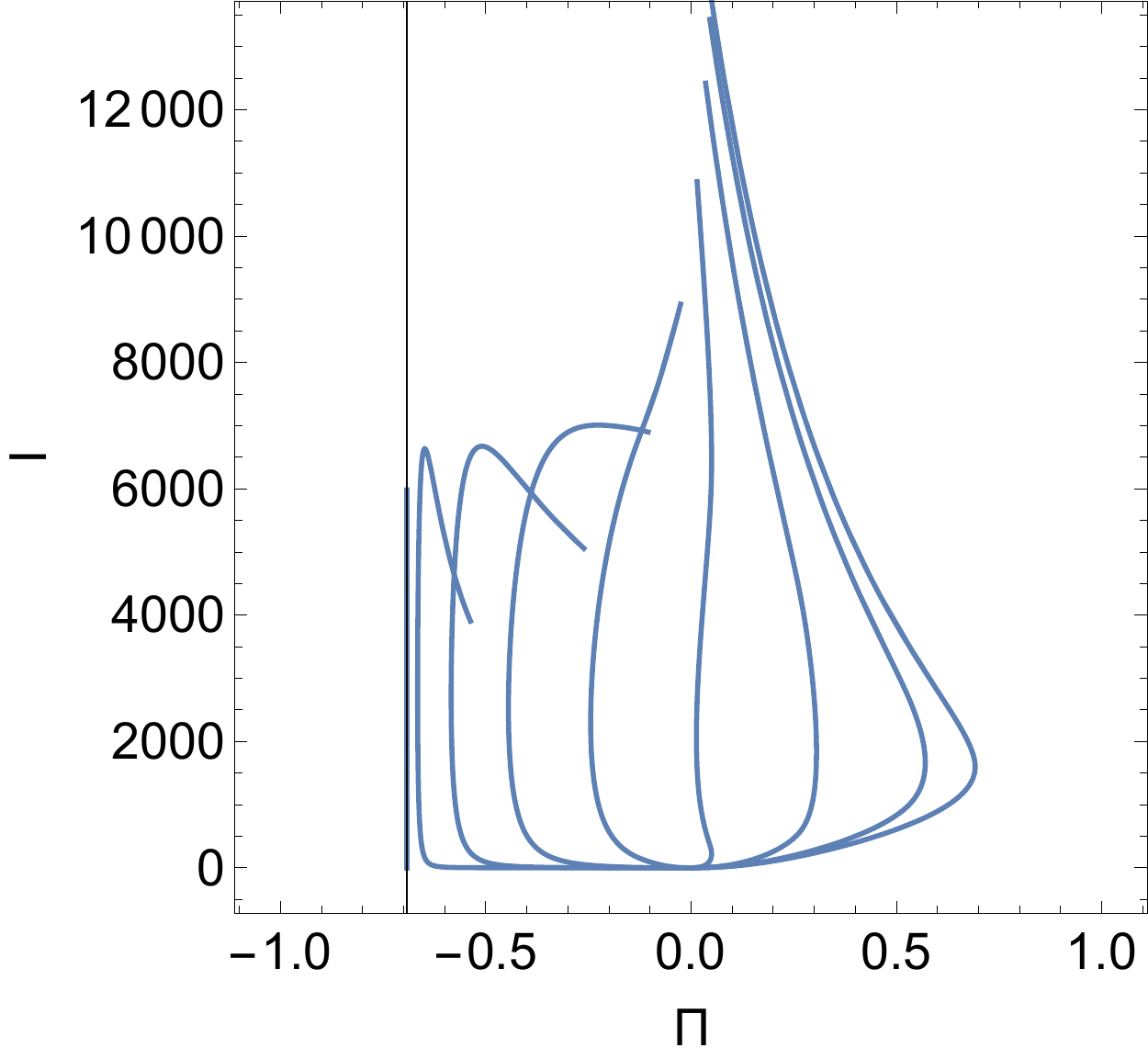}
 \includegraphics[width=.6\columnwidth]{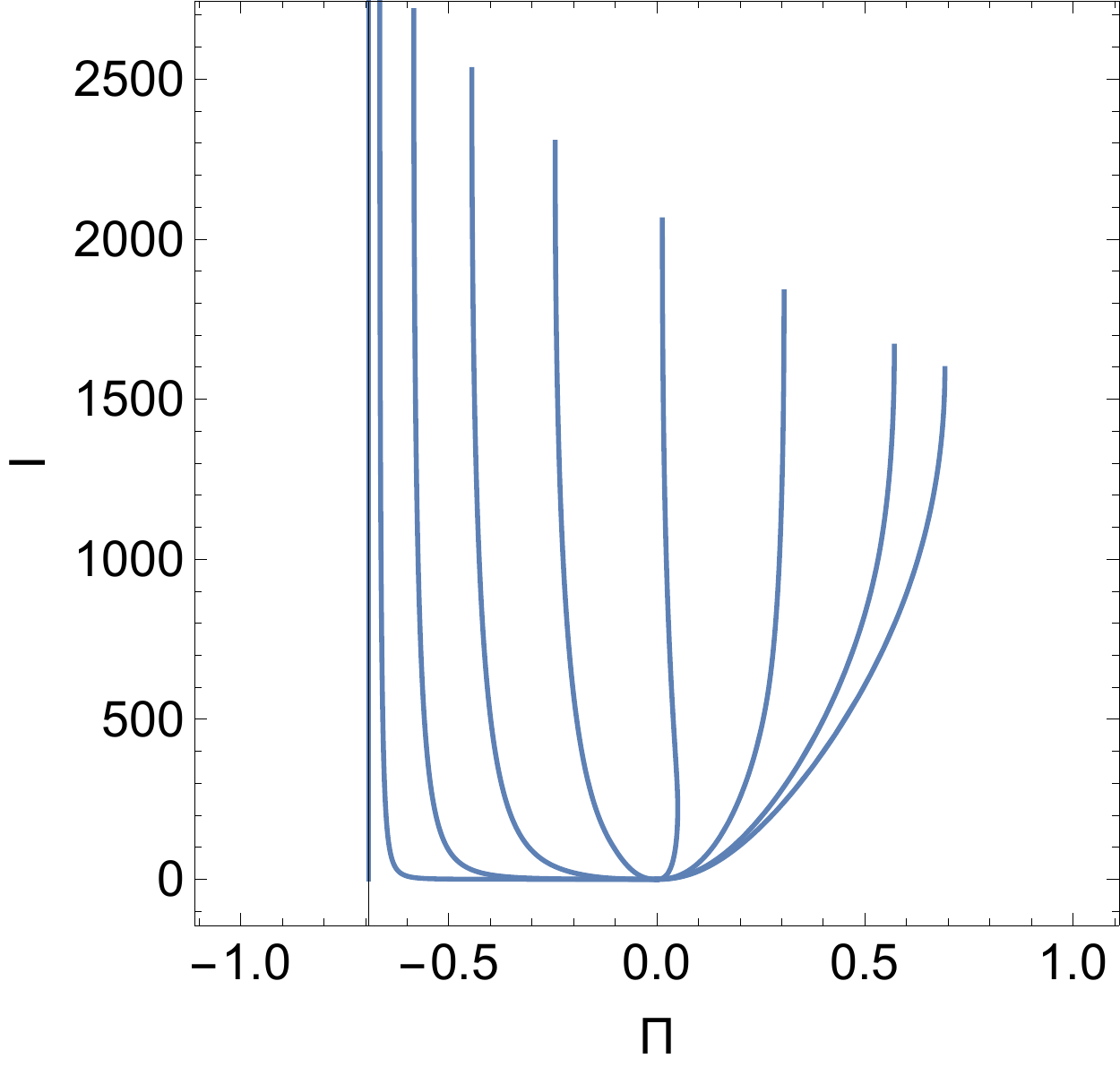}
 \includegraphics[width=.6\columnwidth]{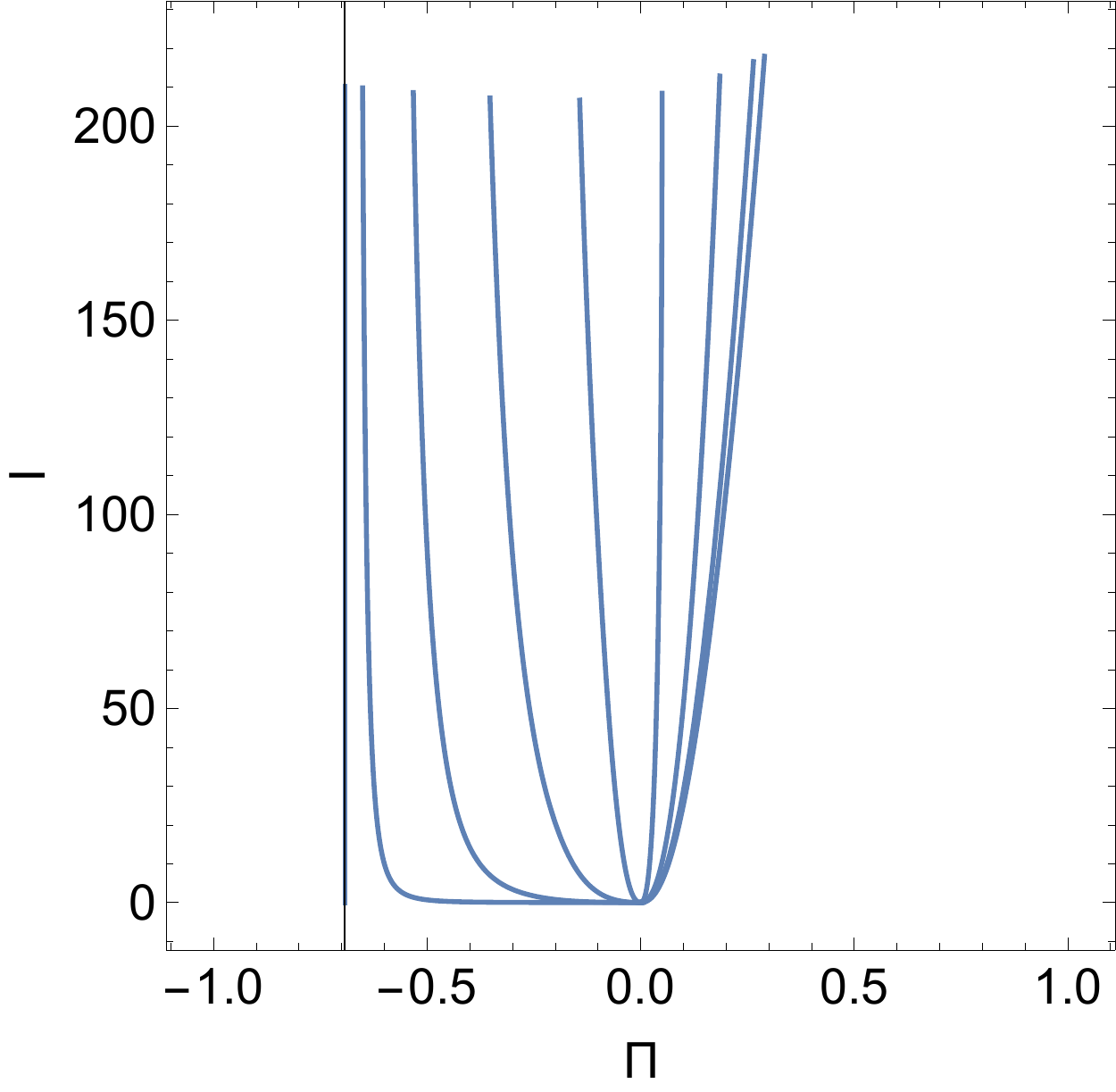}\\
\caption{Same set-up as Fig. \ref{Piofphij}. \textit{Top row}: polarization  $\Pi$ as function of PA. \textit{Bottom Row}: Intensity as function of  polarization degree $\Pi$).} 
\label{Piofphij10}
\end{figure*}
%\end{comment}

Thus, the fastest rate of EVPA swing occurs at $\phi_j =0$. (This reasoning \textit{excludes} possible  fast $\uppi/2$ polarization jumps associated with transitions through $\Pi=0$, see below.)  
Also, the rate of EVPA swing (\ref{phiPA}) is expressed in terms of the coordinate time (and the coordinate rate $ \dot{\phi}_j$). In terms of the observer time these rates will be modified by the time-of-travel effects. In the present paper we concentrate on the overall properties of intensity and polarization and neglect these effects. They will be addressed in a forthcoming paper.

\begin{figure*}
\centering 
$\theta_{ob,0}=1/(5 \gamma), \hskip .15 \columnwidth  \theta_{ob,0}=1/\gamma,  \hskip .15 \columnwidth  \theta_{ob,0}=\, 2/\gamma$\\
 \includegraphics[width=.6\columnwidth]{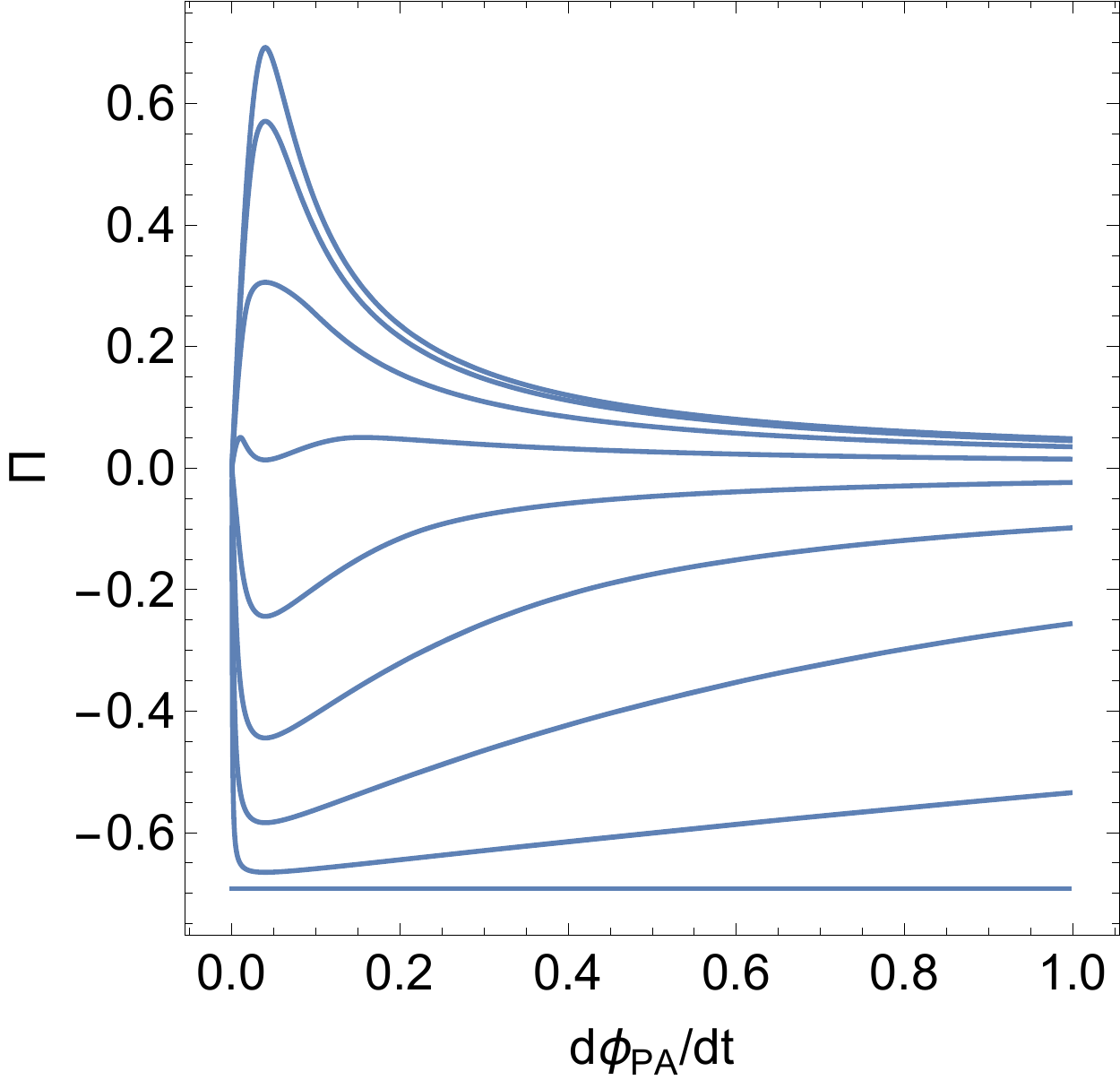}
 \includegraphics[width=.6\columnwidth]{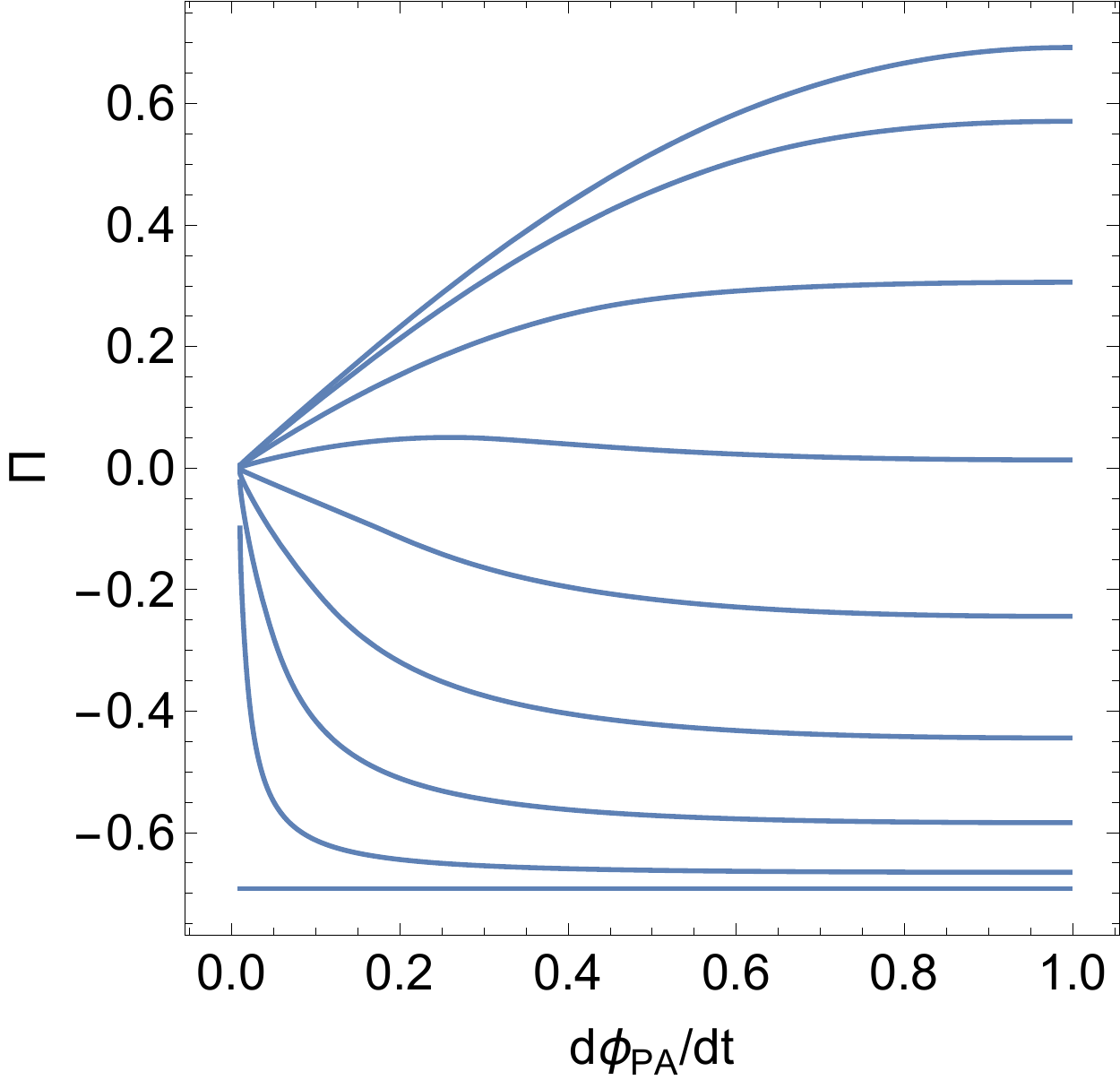}
 \includegraphics[width=.6\columnwidth]{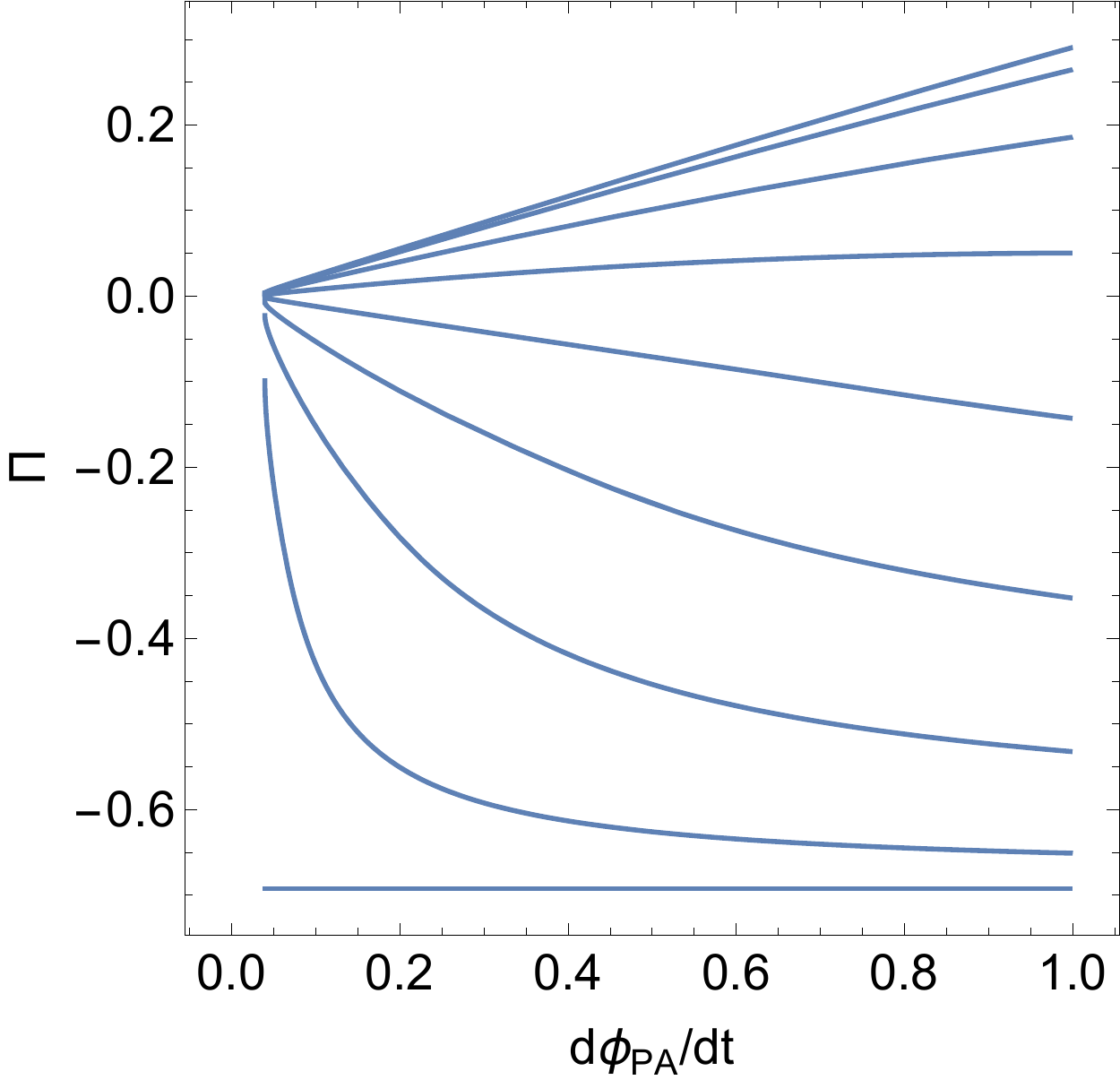}\\
 \includegraphics[width=.6\columnwidth]{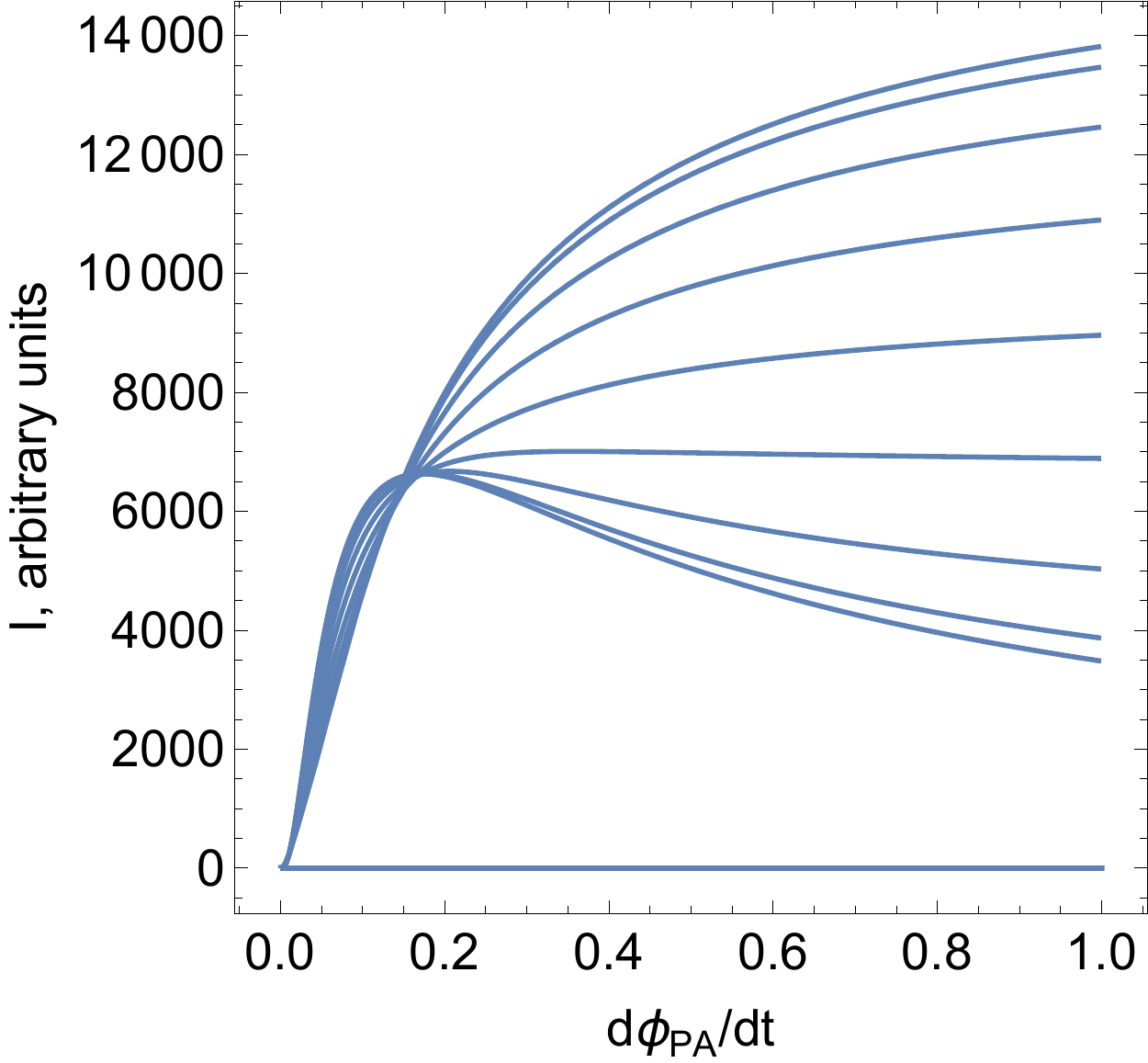}
 \includegraphics[width=.6\columnwidth]{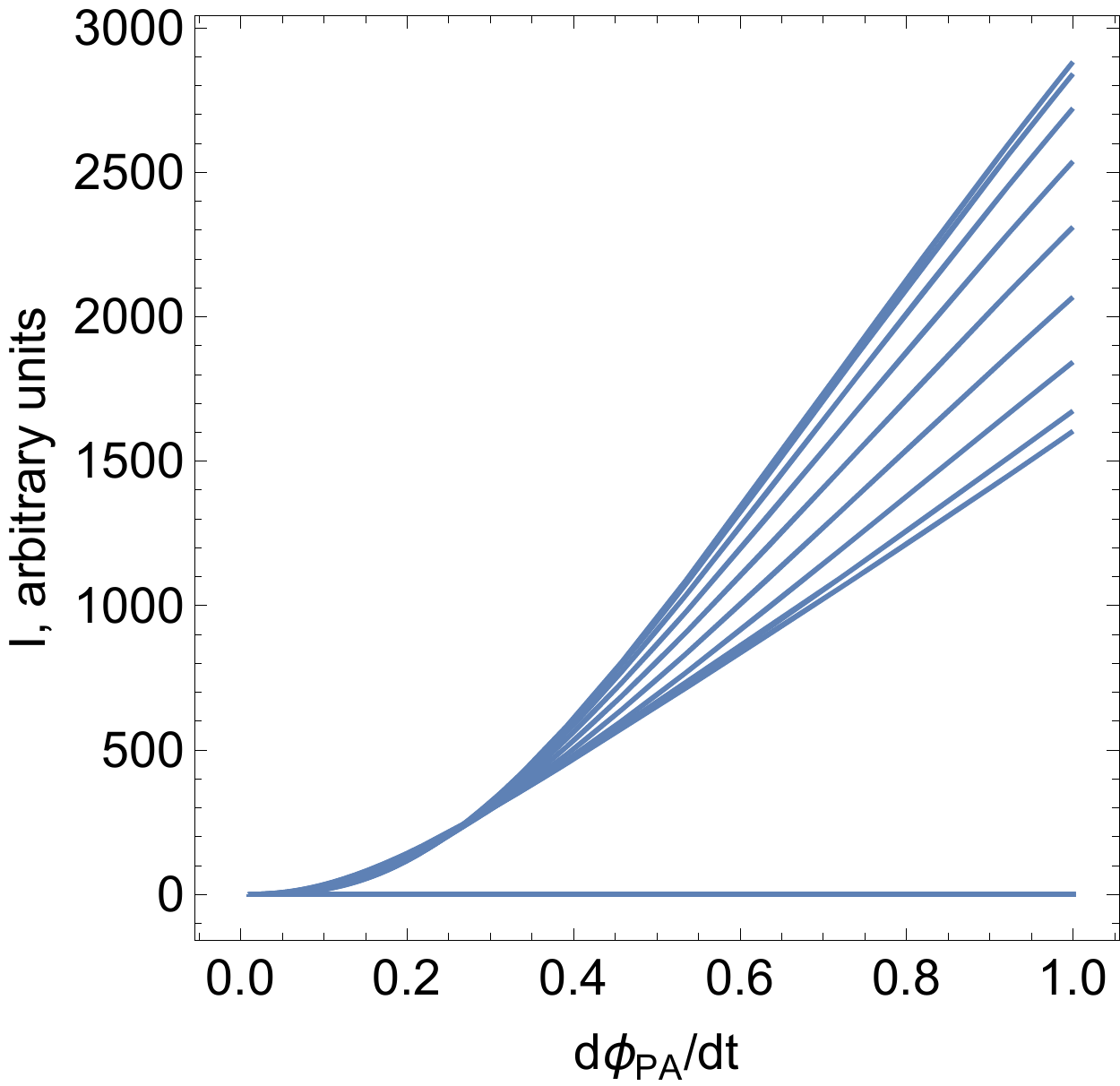}
 \includegraphics[width=.6\columnwidth]{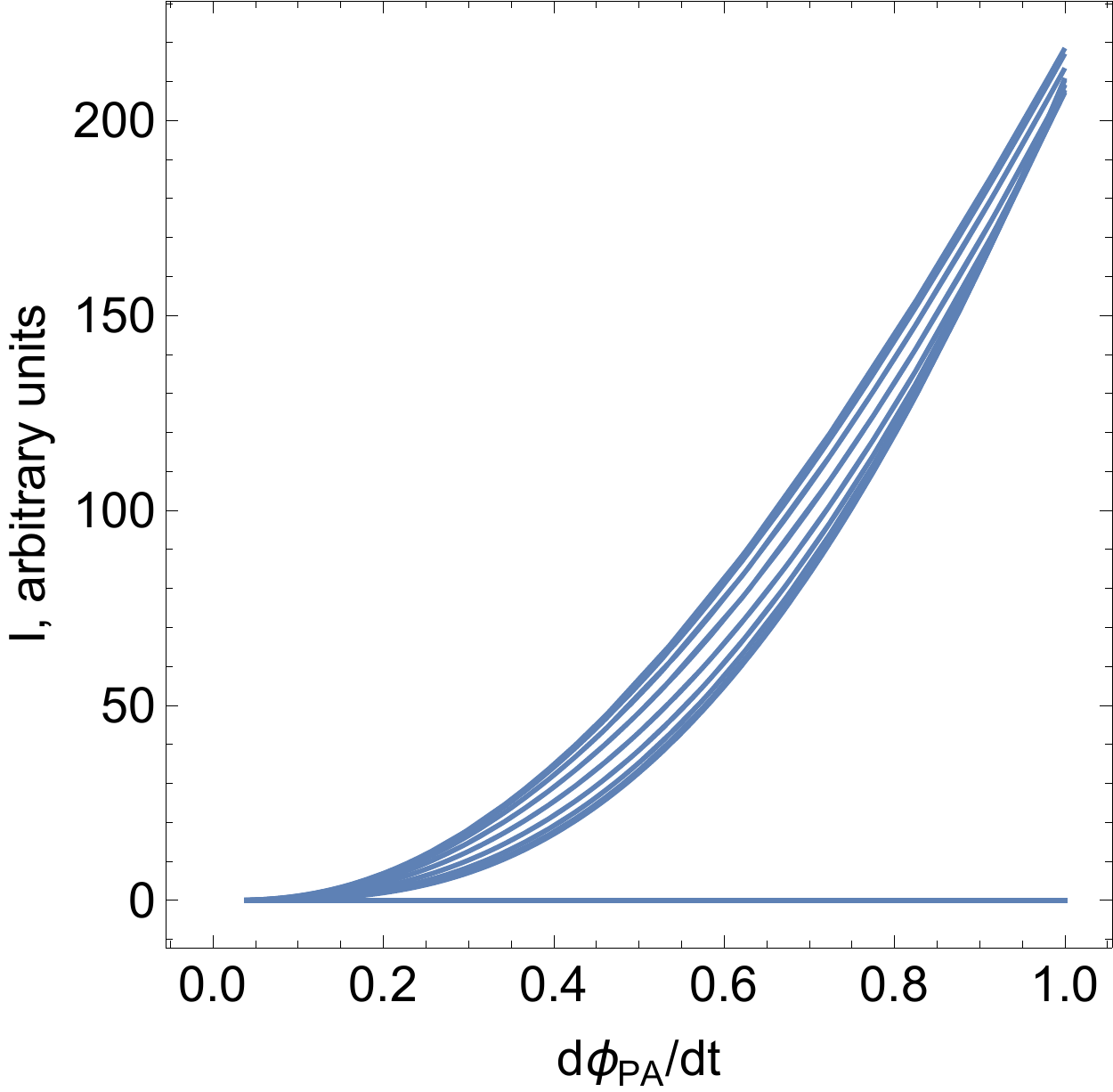}\\
\caption{Same as \ref{Piofphij}. \textit{Top row:} Polarization  $\Pi$ as function of $\dot{\phi}_{PA}$. Polarization can be both minimal at fastest EVPA swings (left panel) or maximal (center and right panels). For some values of the pitch angle the polarization fraction is a non-monotonic function of $\dot{\phi}_{PA}$. \textit{Bottom row:} Intensity as function of $\dot{\phi}_{PA}$. Intensity generally increases with  $\dot{\phi}_{PA}$, but can be non-monotonic (left panel). (The rate $\dot{\phi}_{PA}$ is normalized to  a maximal value).
} 
\label{Piofphij1}
\end{figure*}

\begin{figure}
\centering 
\includegraphics[width=.6\columnwidth]{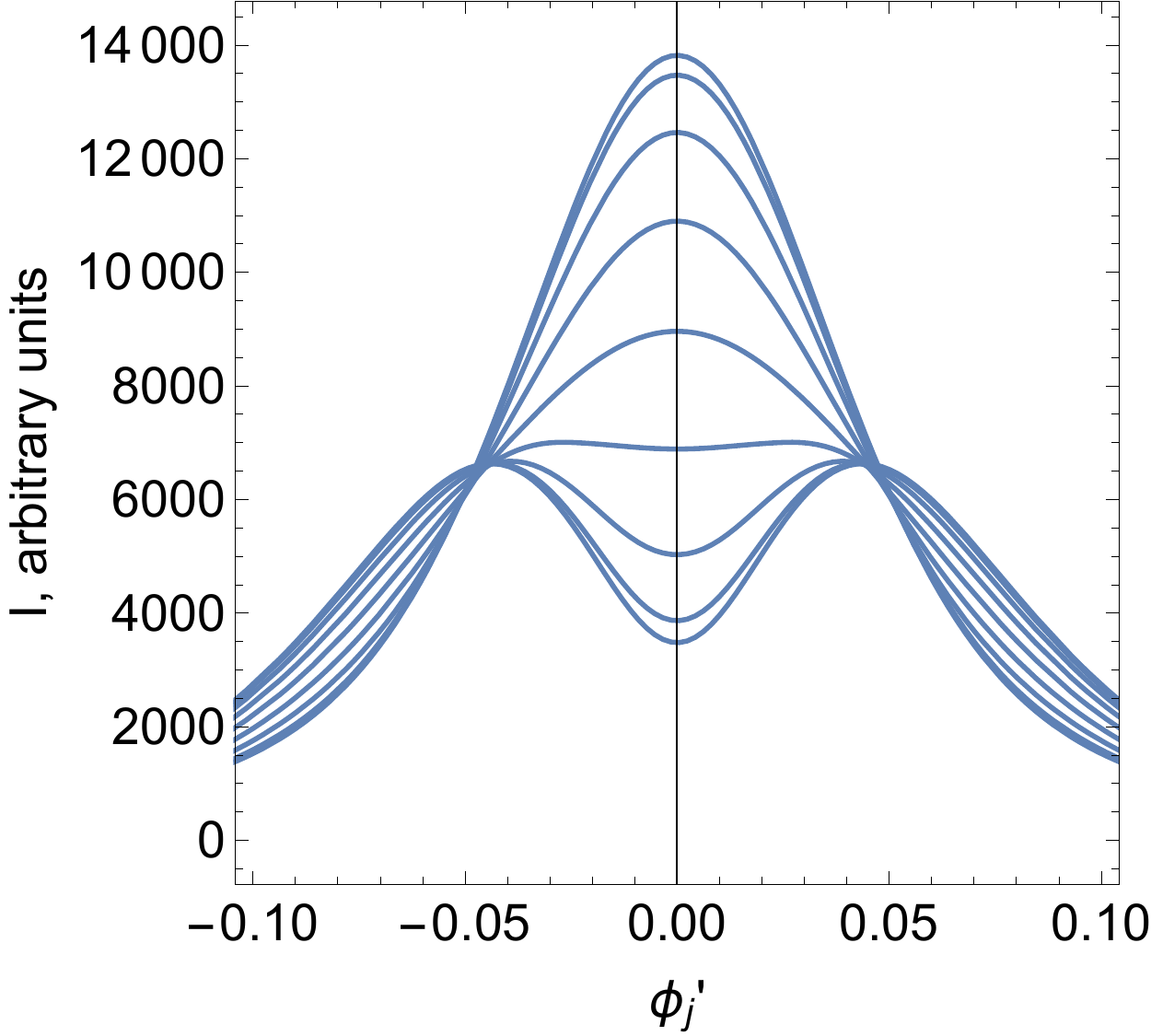}\\
\includegraphics[width=.6\columnwidth]{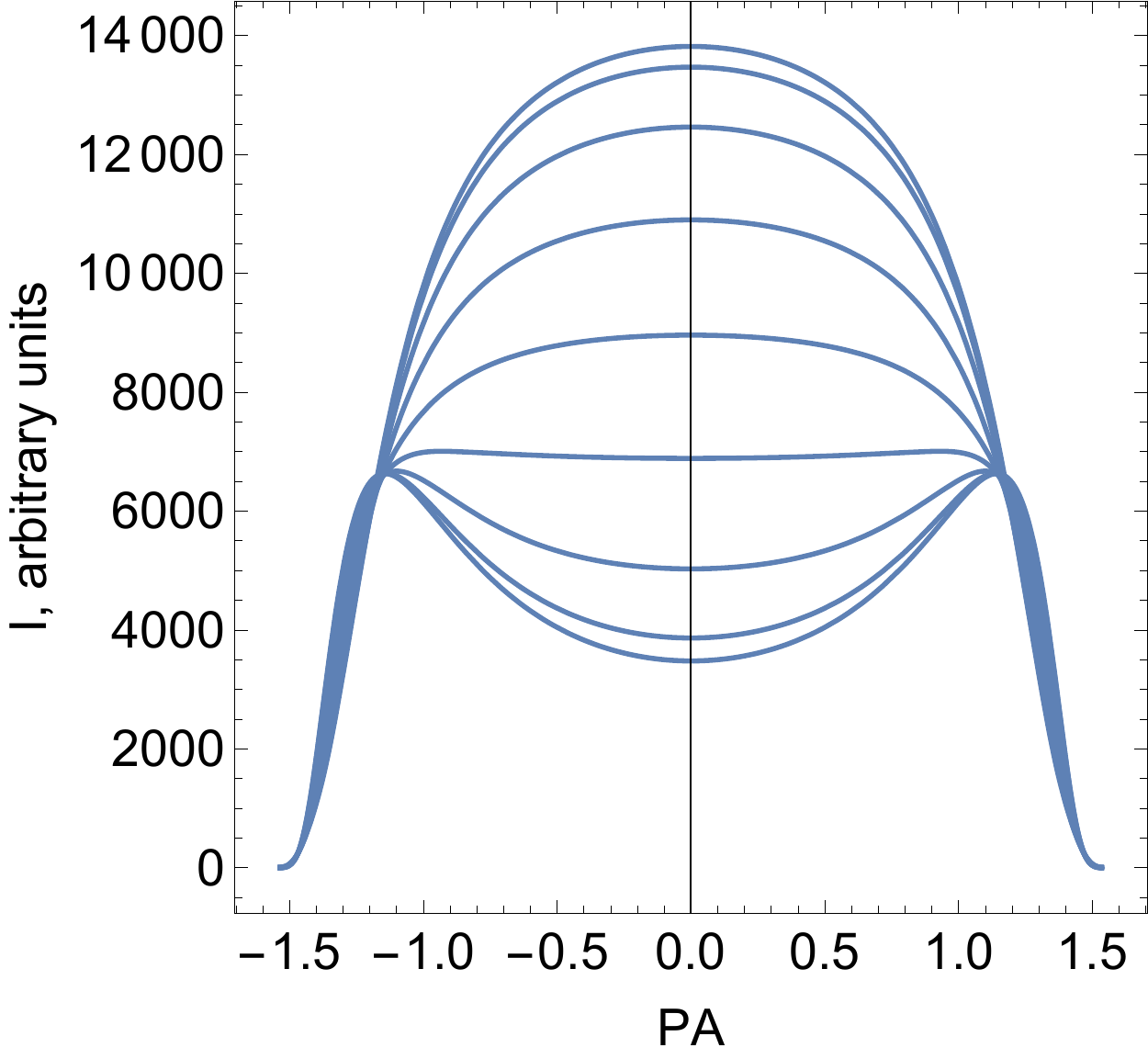}
\caption{ \textit{Top panel}: Intensity as function of the oscillation angle (only limited range in $\phi_j$ is shown).  For small pitch angles, and small $\theta_{ob,0}$ intensity in the middle can be low (looking almost along   the \Bf).  \textit{Bottom  panel}:  Intensity as function of EVPA.} 
\label{PiofphiII2}
\end{figure}

In Figs. \ref{Piofphij}-\ref{Piofphij10}-\ref{Piofphij1}, we plot the  polarization signatures assuming that the motion of the jet is symmetric with respect to the line of sight and that oscillations occur between angles $-\uppi/2 < \phi_j < \uppi/2$. 
{We note, for $\Pi>0$ the polarization is along the jet, while the polarization is orthogonal to the jet for $\Pi<0$.}
In Fig. \ref{PiofphiII2} we plot the observed intensity as a function of the oscillation angle and as a function of the  EVPA of polarization.

%\begin{comment}

%\begin{comment}
%\end{comment}

\begin{figure}
\centering 
\includegraphics[width=.67\columnwidth]{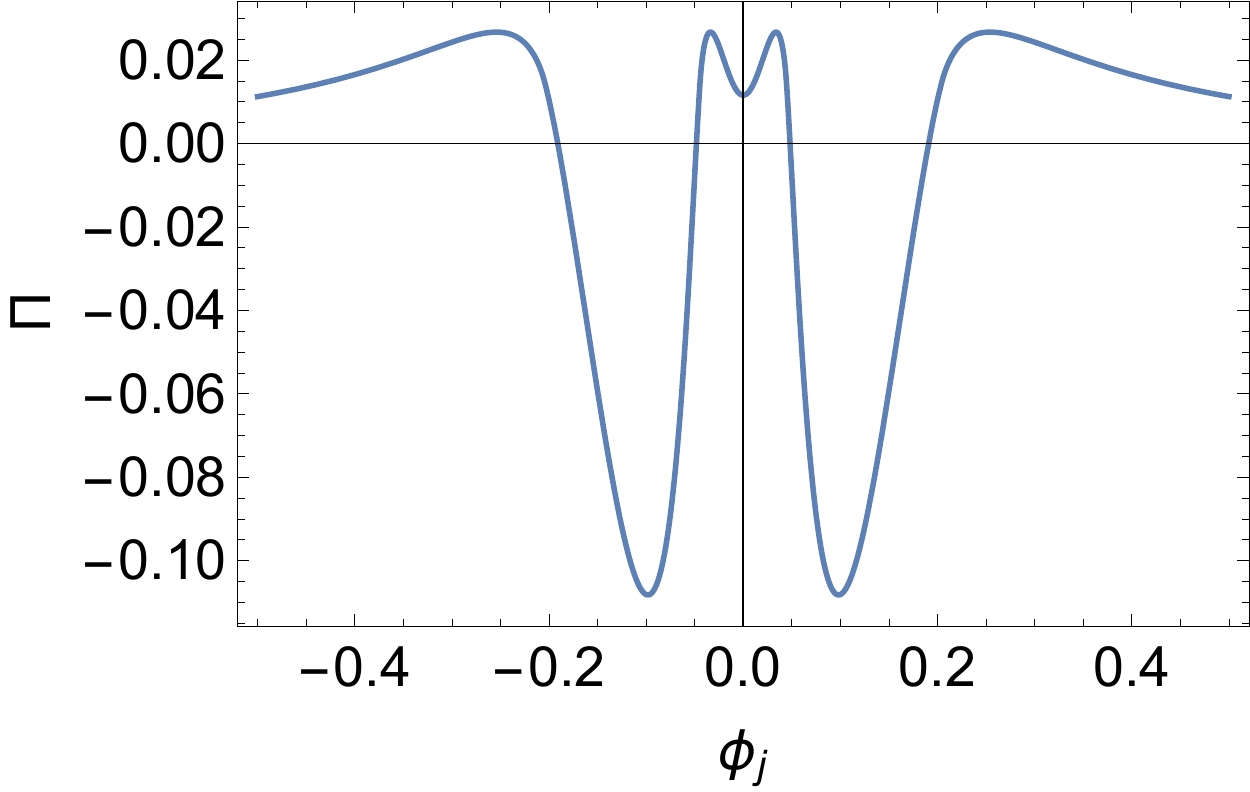}\\
\includegraphics[width=.6\columnwidth]{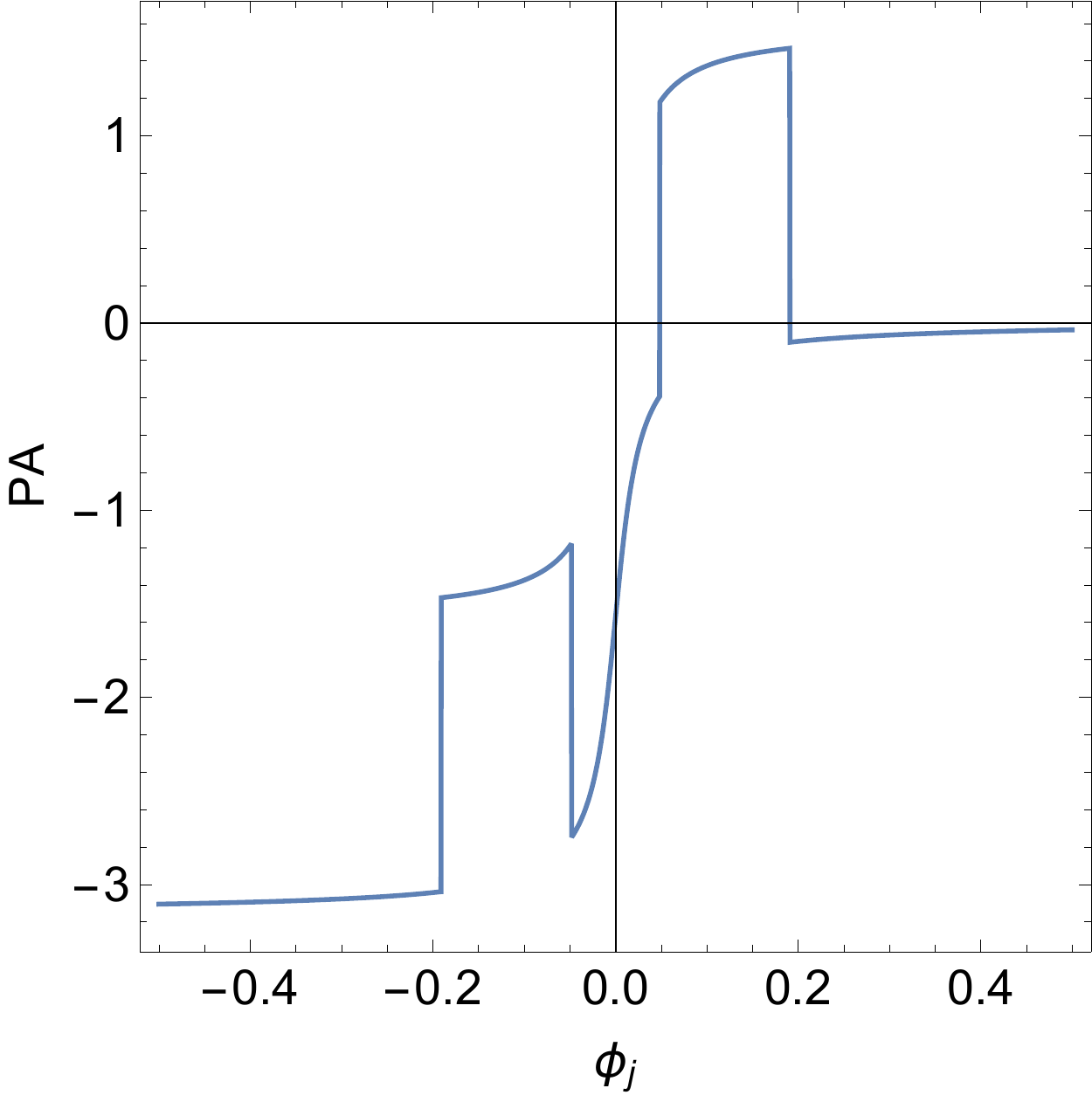}
\caption{Polarization fraction and EVPA for $\psi = 2 \uppi/7$ (the direction of the jumps in EVPA are undetermined up to the factor  $\pm \uppi/2$); at the moment of the jump $\Pi=0$.} 
\label{2piover7}
\end{figure}

There are two types of fast  EVPA variation: (i) when the jet passes close to the line of sight, EVPA experiences fast smooth variations that can approach $\uppi$ radians; (ii) occasionally EVPA experiences sudden jumps by $\uppi/2$ radians (Fig. \ref{2piover7}). During such jumps the polarization fraction passes through zero.

%\begin{comment}
%\end{comment}

\subsection{Accelerating and swinging  jets}
\begin{figure*}
\centering 
\includegraphics[width=.7\columnwidth]{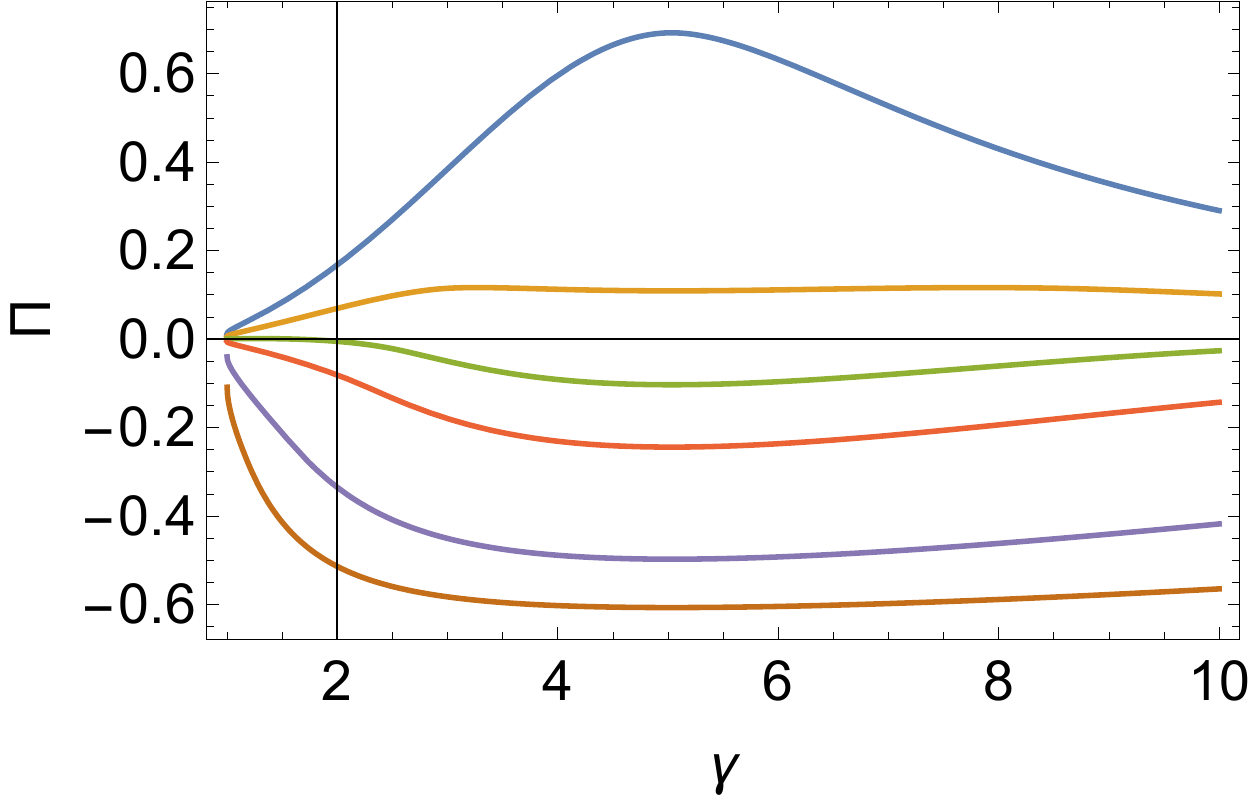}\quad
\includegraphics[width=.7\columnwidth]{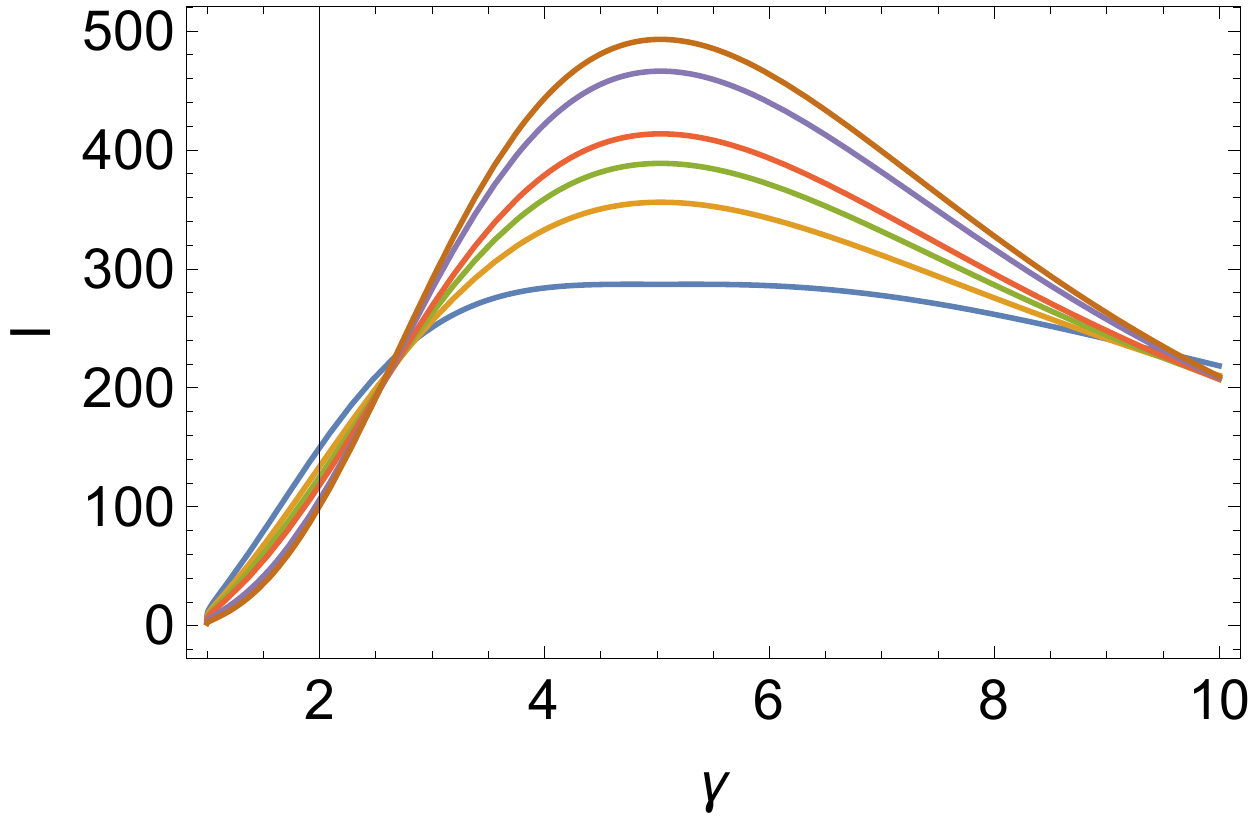}\quad
\includegraphics[width=.55\columnwidth]{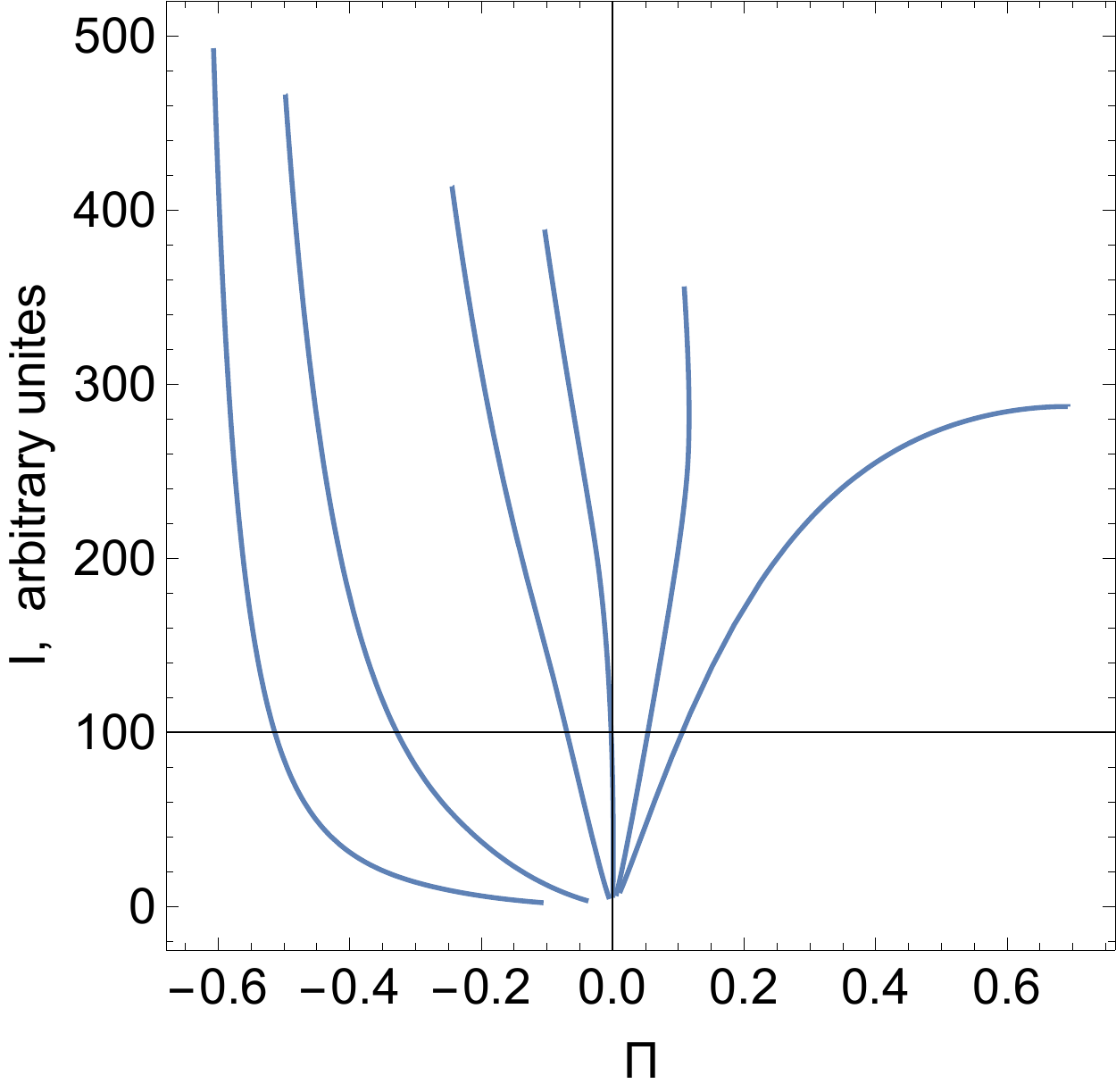}\\
\caption{Polarization $\Pi$ (left panel) and intensity $I$ (center panel)  as function of bulk \Lf\  $\gamma$ for accelerating jet.  Intensity as function of polarization (right panel).   Viewing angle $\theta_{ob} = 0.2$. Different curves correspond to different intrinsic pitch angles.} 
\label{Piofgamma}
\end{figure*}

Next we consider effects of jet acceleration. In Fig. \ref{Piofgamma} we plot polarization and intensity as function of the bulk \Lf\ (left and center panels) and intensity as function of polarization fraction (right panel).
Naturally, intensity is maximal when $\gamma \sim 1/\theta_{ob}$ - for smaller $\gamma$ the effects of relativistic boosting are small, while for higher $\gamma$ the emission is boosted away from the observer.
The polarization fraction is typically smaller at lower intensities, but shows a variety of behavior. 

%\begin{comment}
%\end{comment}

\begin{figure}
\centering 
\includegraphics[width=.6\columnwidth]{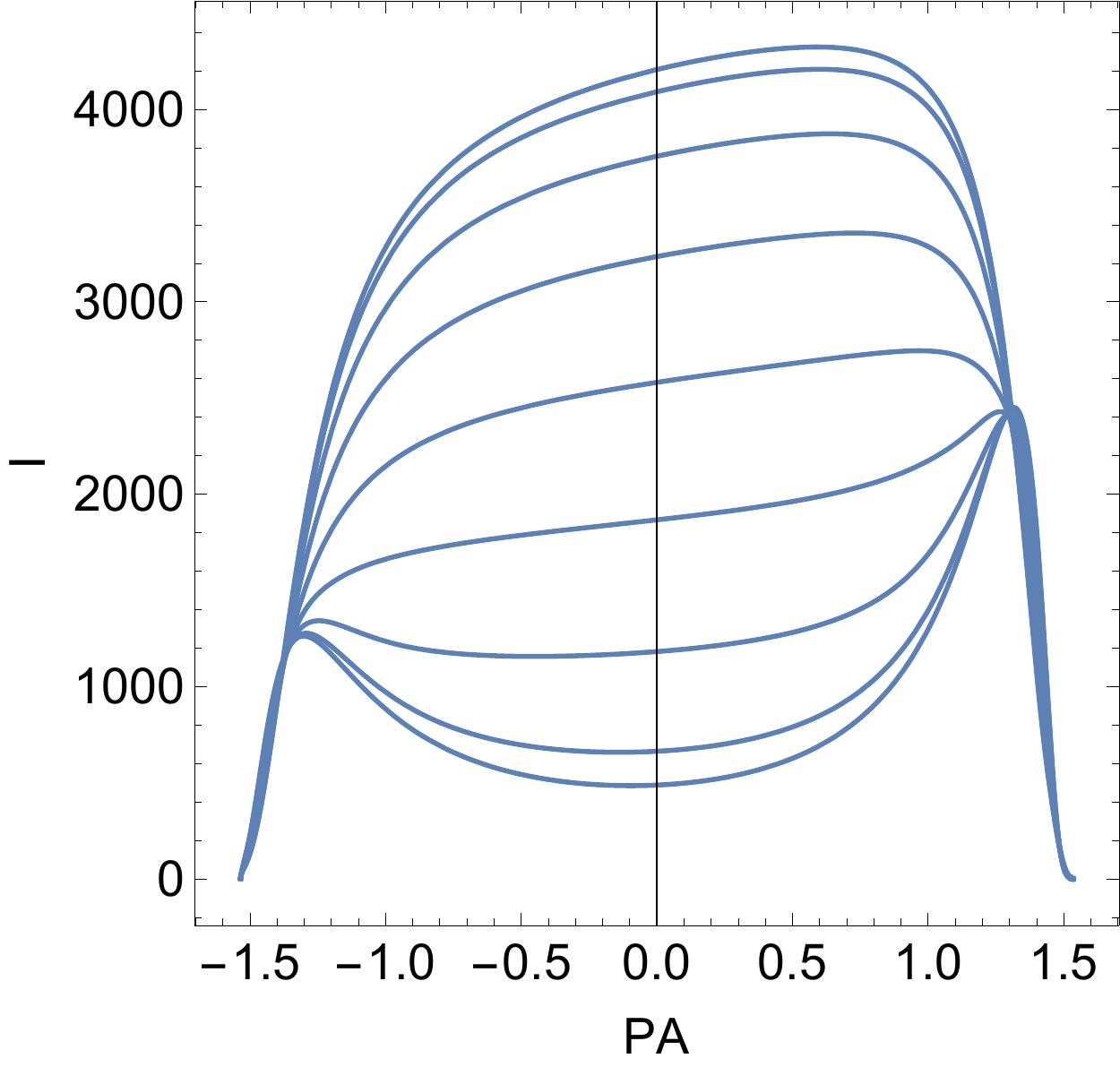}\\
\includegraphics[width=.6\columnwidth]{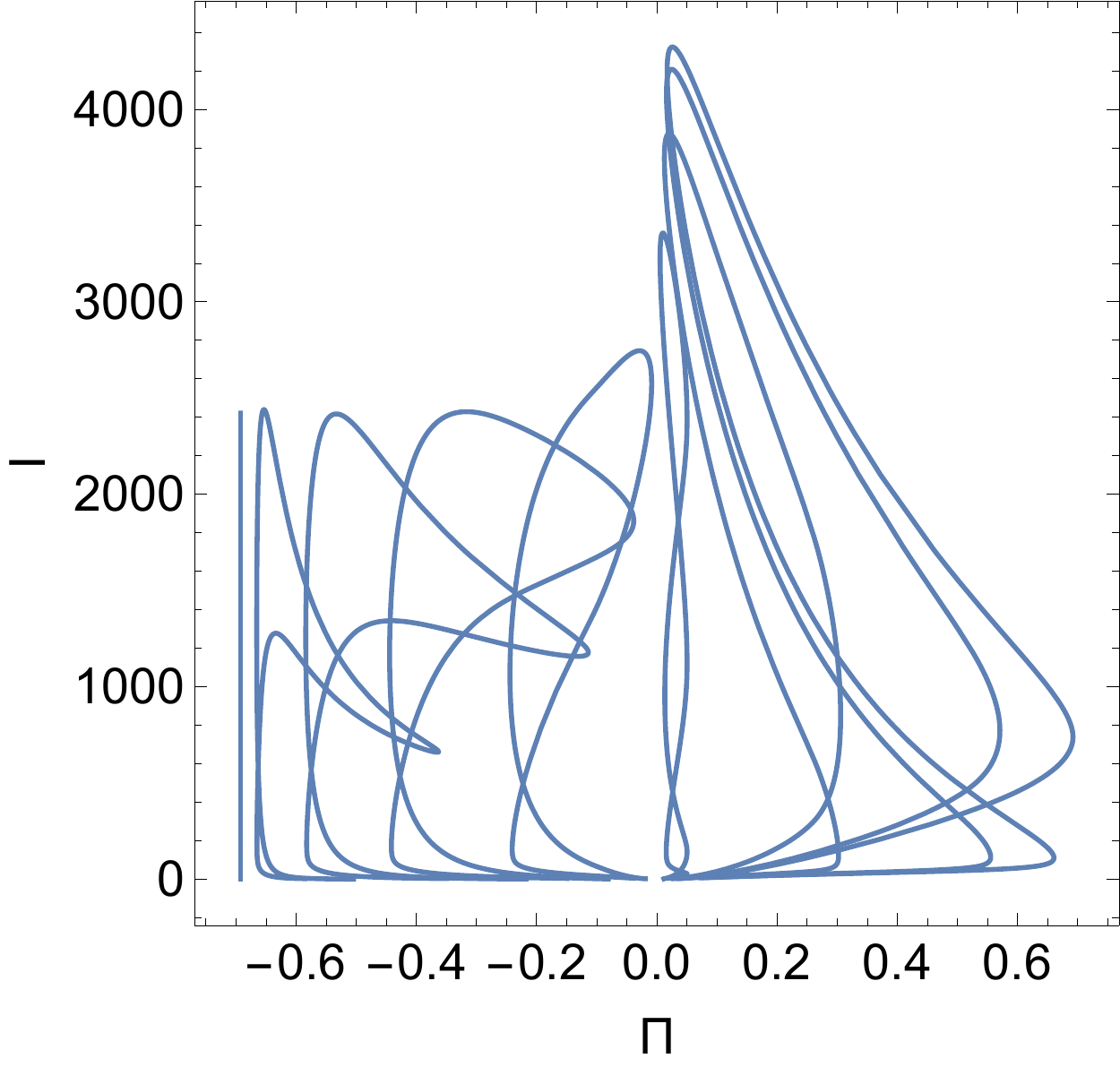}
\caption{Swinging and accelerating jets. Shown are intensity as function of the  EVPA (top panel) and polarization fraction (bottom panel).  Though in most cases peak intensity coincides with minimal $\Pi$, there are clear exceptions, when polarization can be maximal at the peak of intensity, or be flat. } 
\label{PiofgammaS}
\end{figure}

Next we consider swinging and accelerating jets, Fig. \ref{PiofgammaS}. The jet executes a planar motion with amplitude $- 5/\gamma_{max}< \phi < 5/\gamma_{max}$, $ \gamma_{max}=10$, while the \Lf\ increases linearly from $\gamma=1$ at $\phi = - 5/\gamma_{max}$ to $\gamma=\gamma_{max}$ at $\phi =  5/\gamma_{max}$. The minimal viewing angle is $\theta_{ob, min} = 1/(5 \gamma_{max})$. The general feature that acceleration introduces is a non-symmetric behavior of polarization with respect to the swing of  EVPA.

%\begin{comment}
%\end{comment}

\subsection{Circular motion}
\label{circula}
\begin{figure*}
\centering
\includegraphics[width=.58\columnwidth]{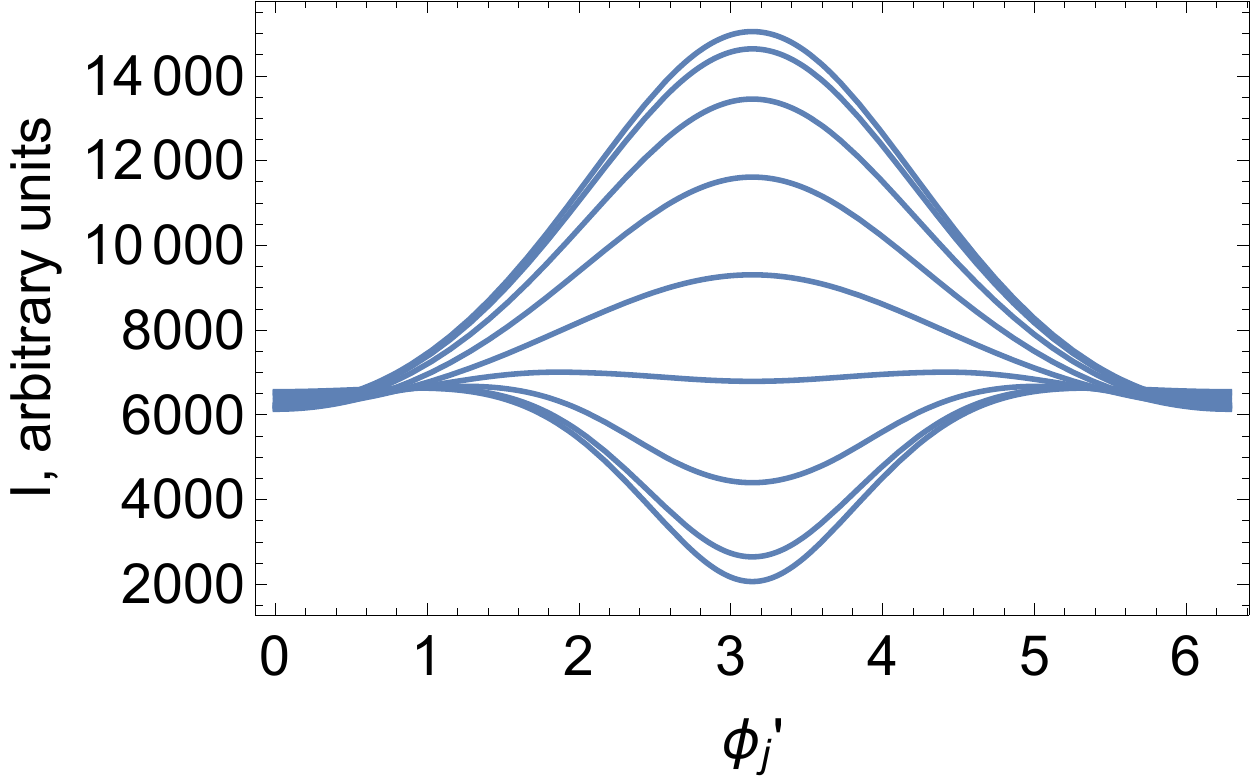}
\includegraphics[width=.58\columnwidth]{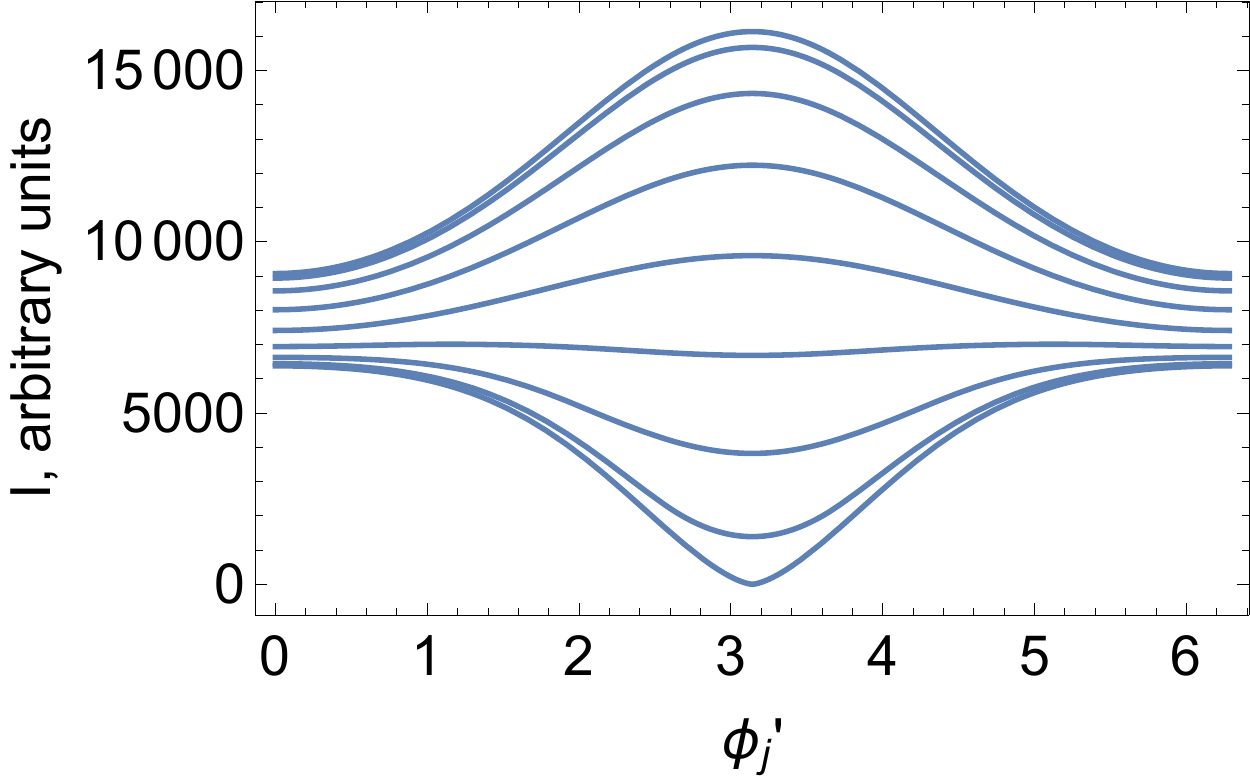}
\includegraphics[width=.58\columnwidth]{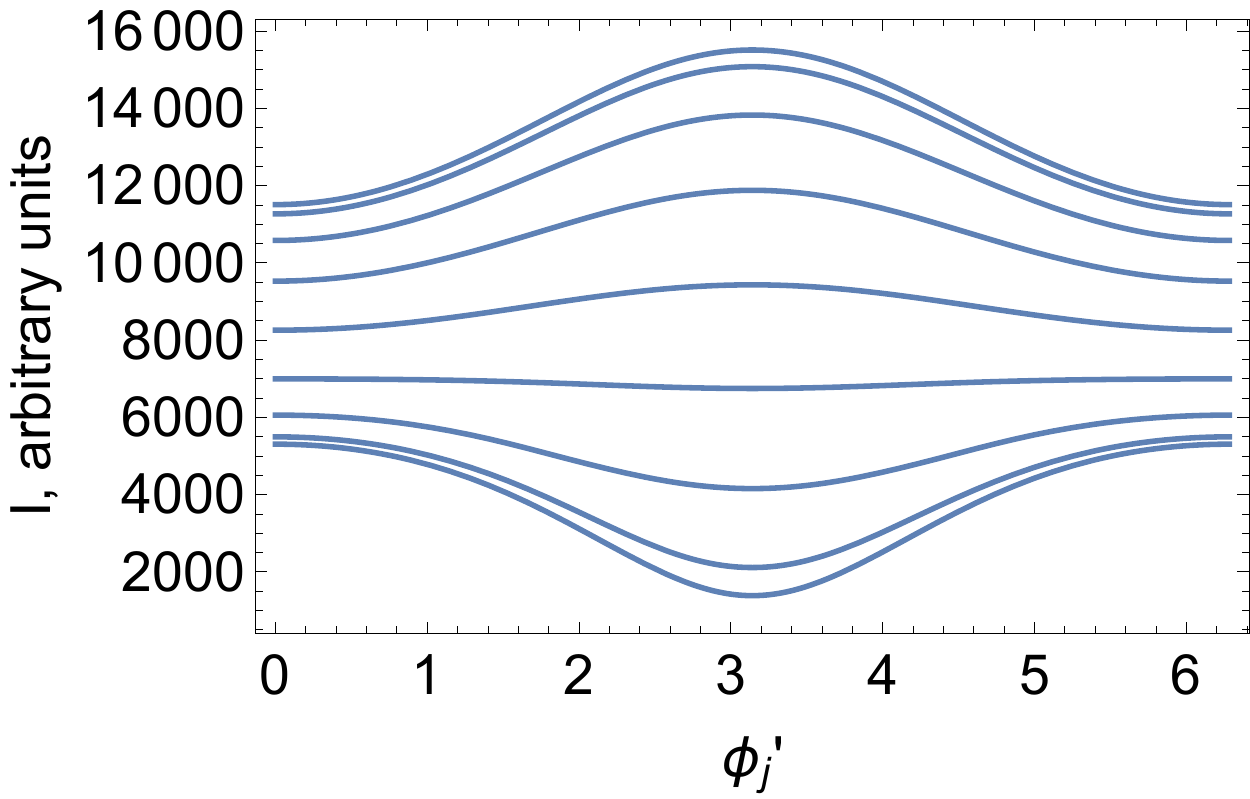}\\
\includegraphics[width=.58\columnwidth]{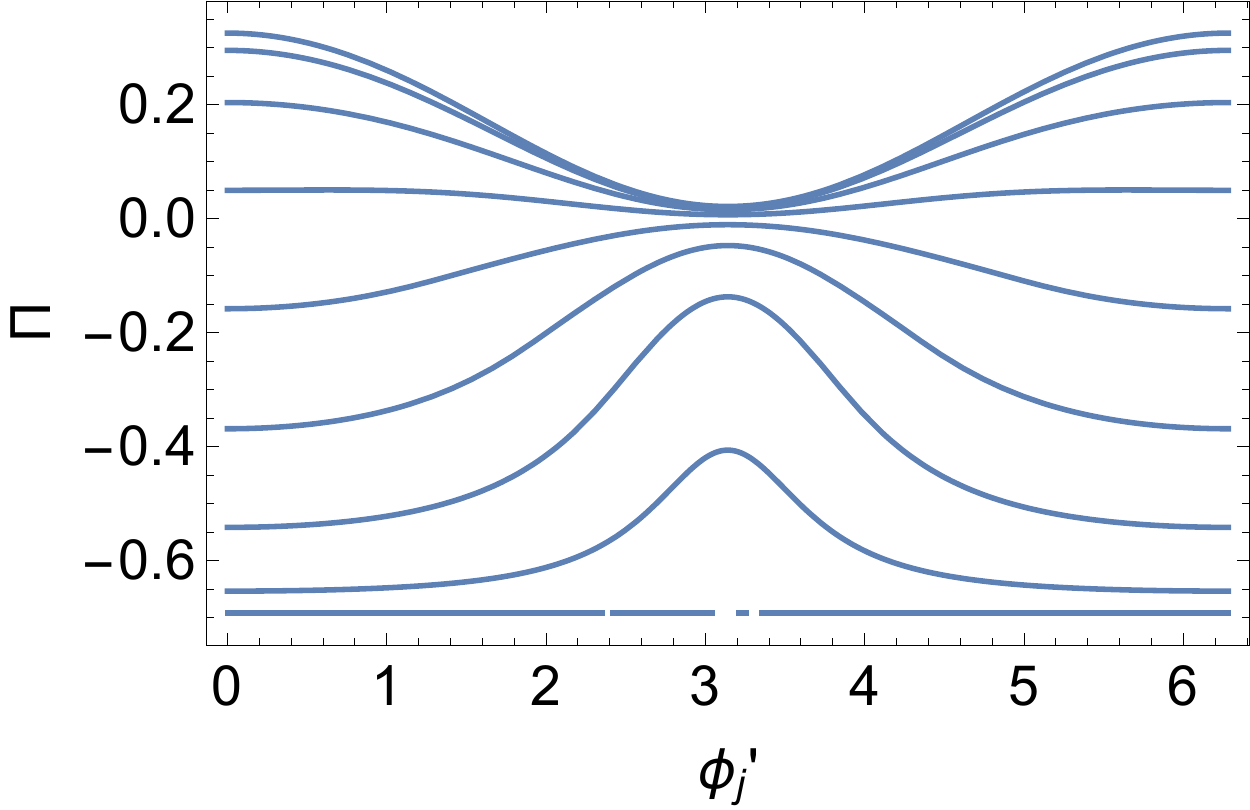}
\includegraphics[width=.58\columnwidth]{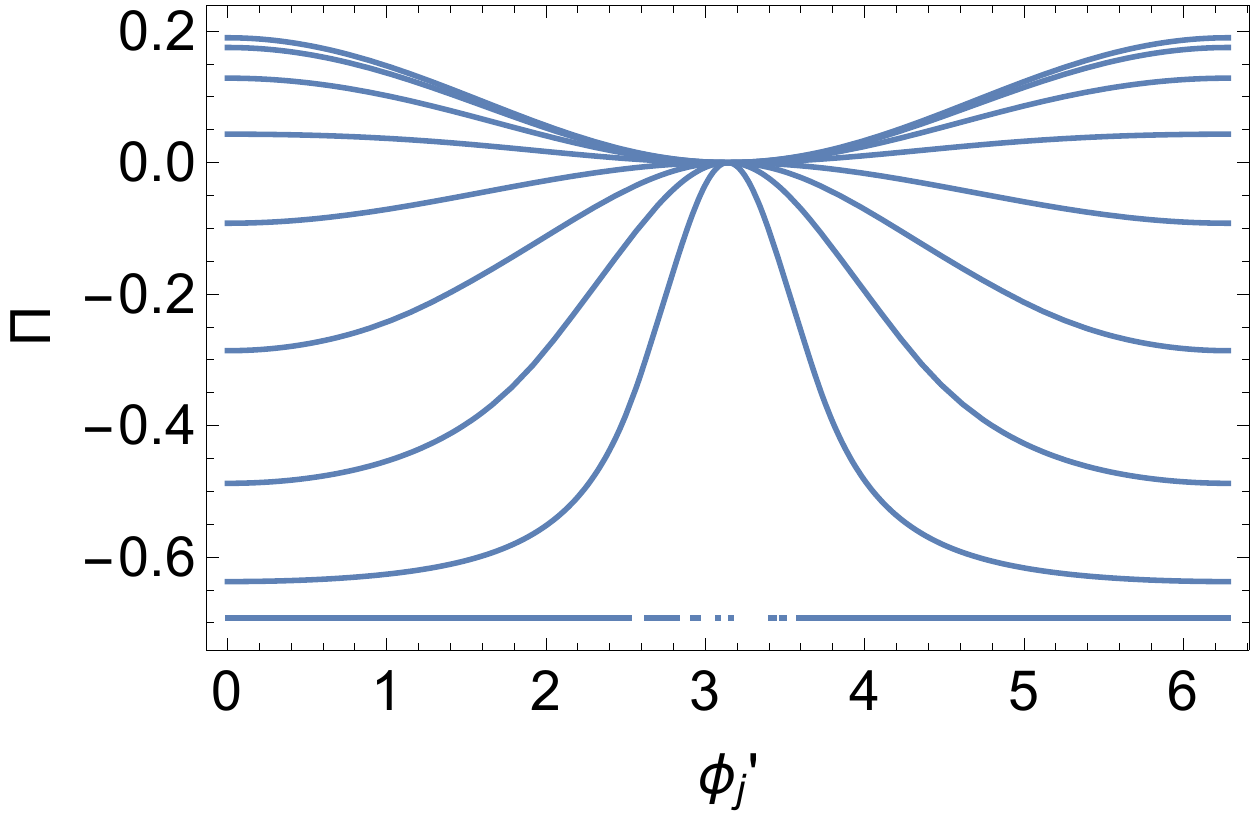}
\includegraphics[width=.58\columnwidth]{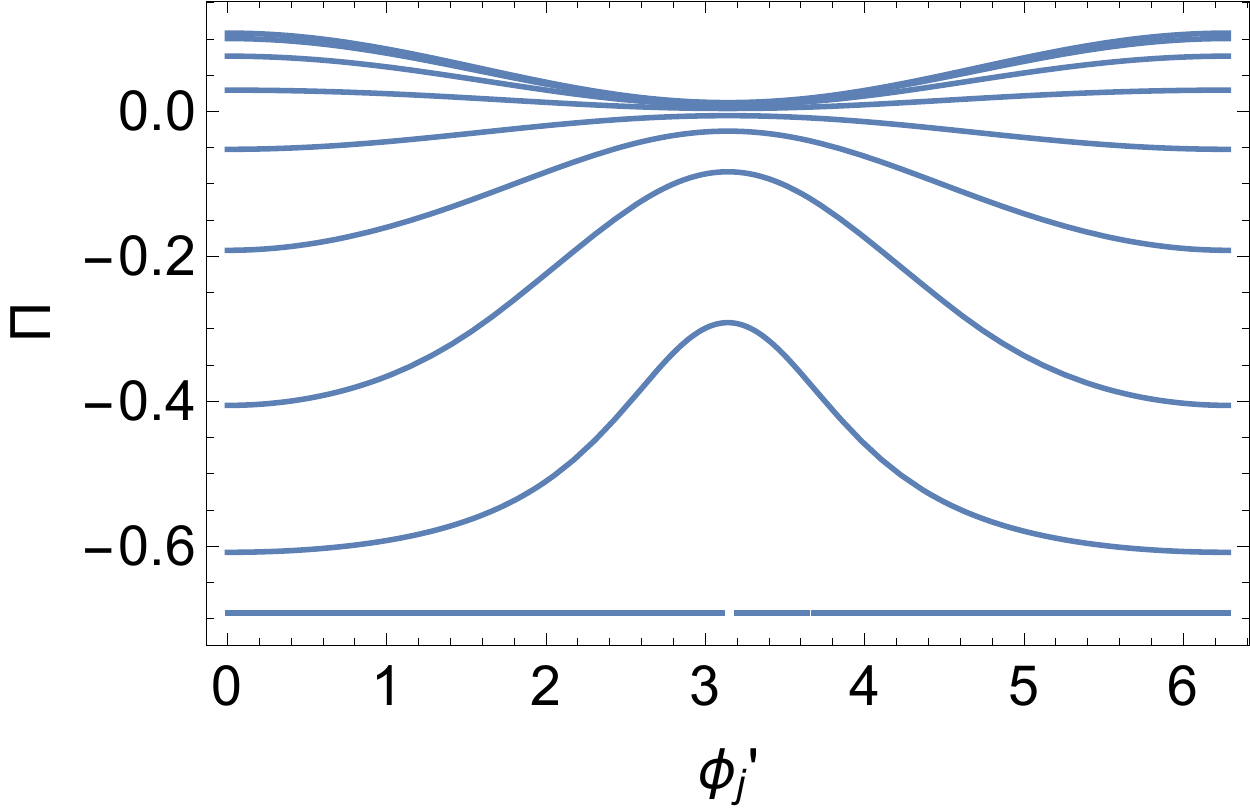}\\
\includegraphics[width=.58\columnwidth]{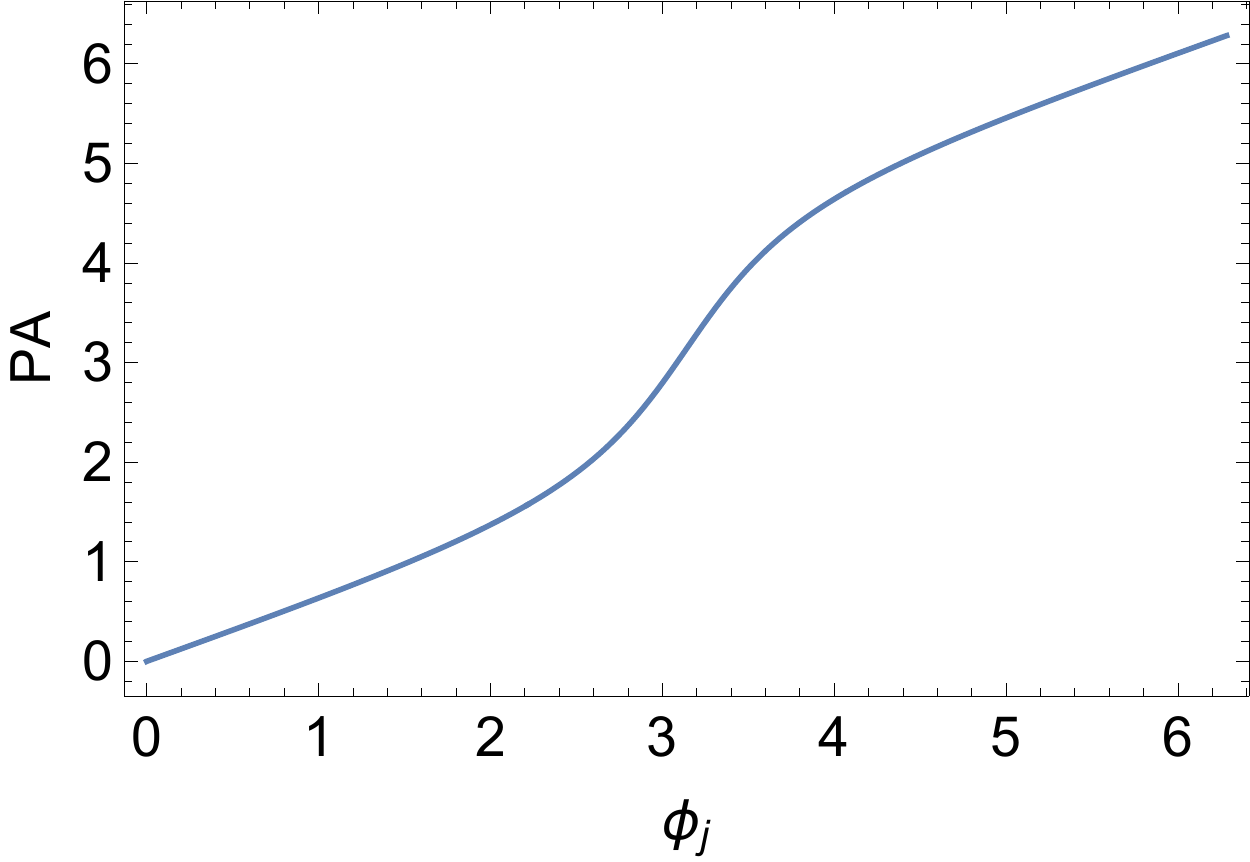}
\includegraphics[width=.58\columnwidth]{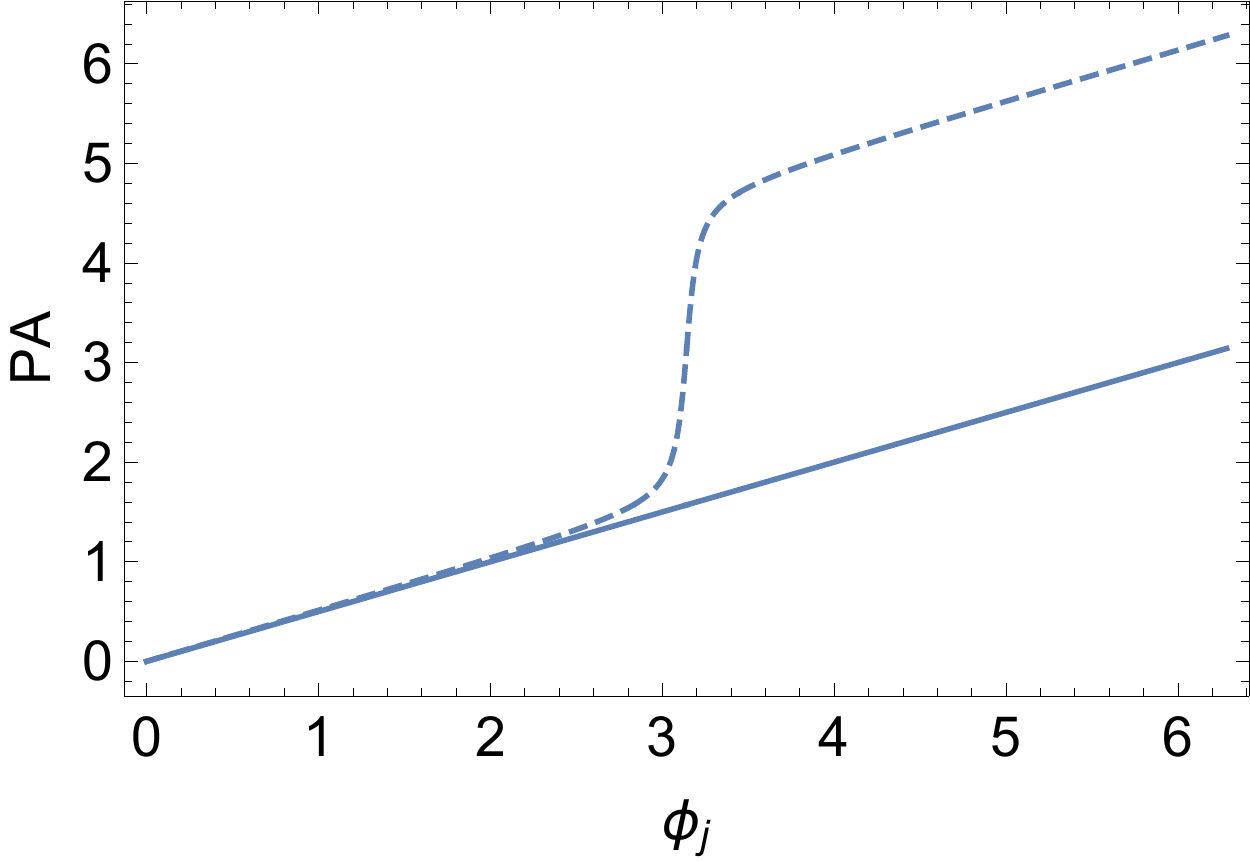}
\includegraphics[width=.58\columnwidth]{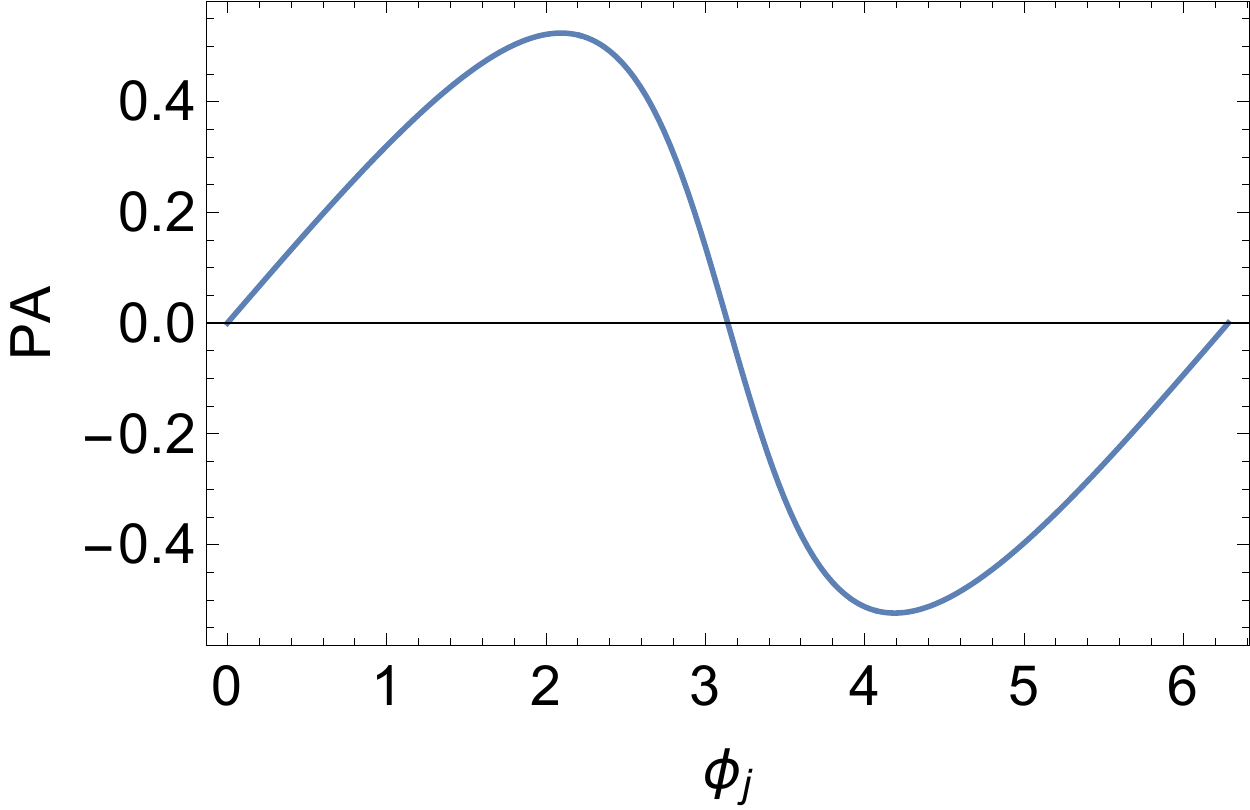}\\
\includegraphics[width=.58\columnwidth]{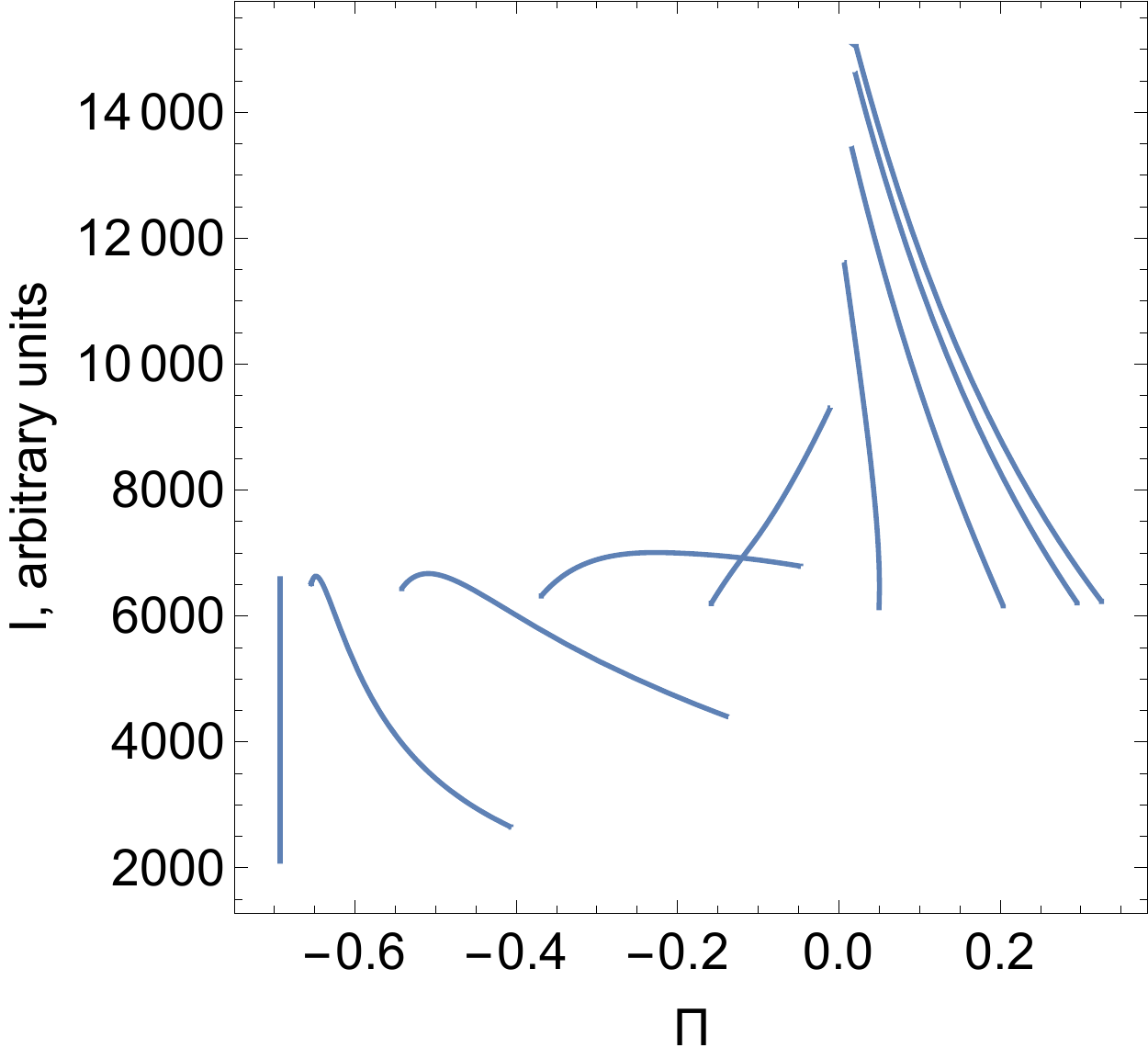}
\includegraphics[width=.58\columnwidth]{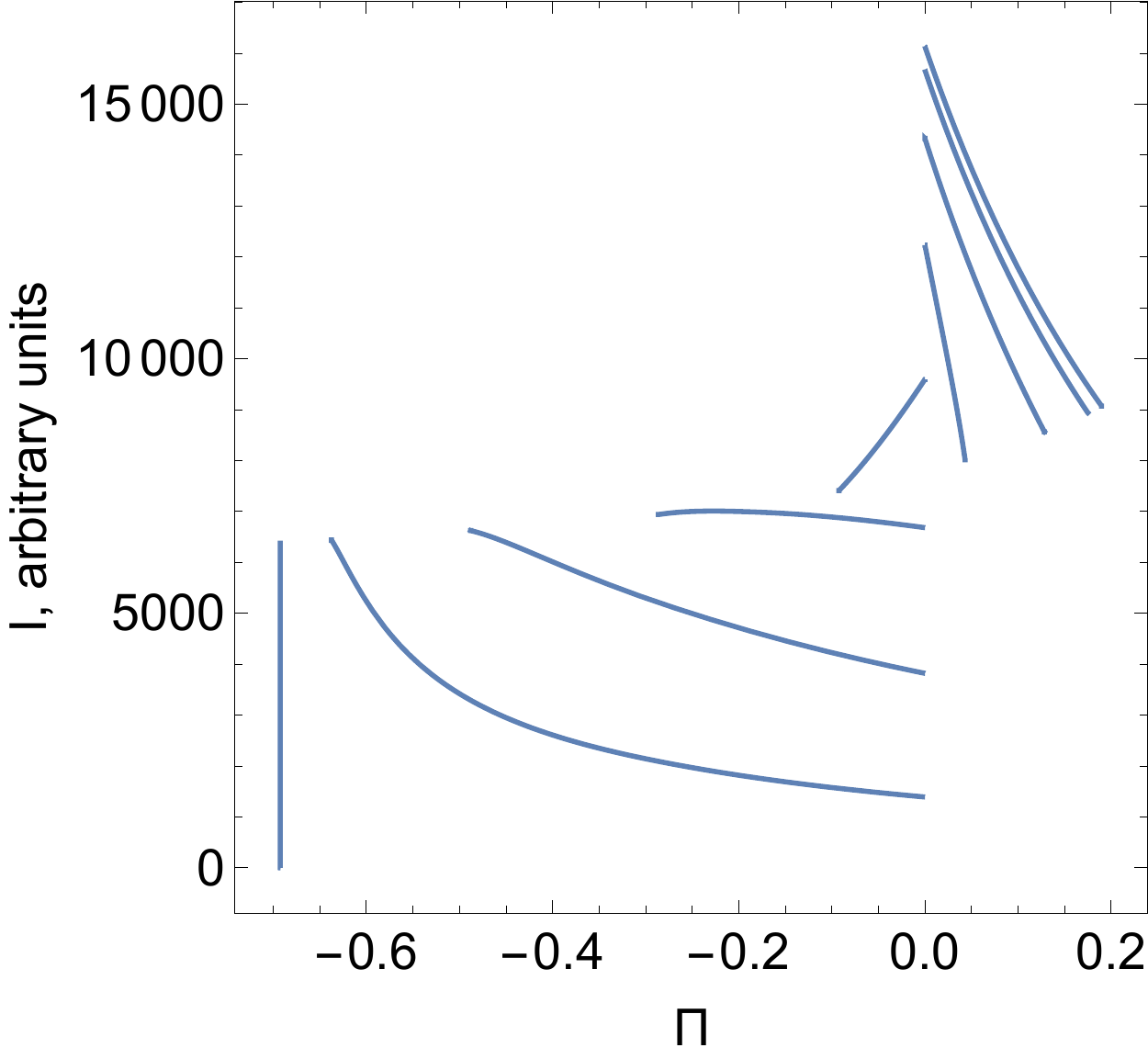}
\includegraphics[width=.58\columnwidth]{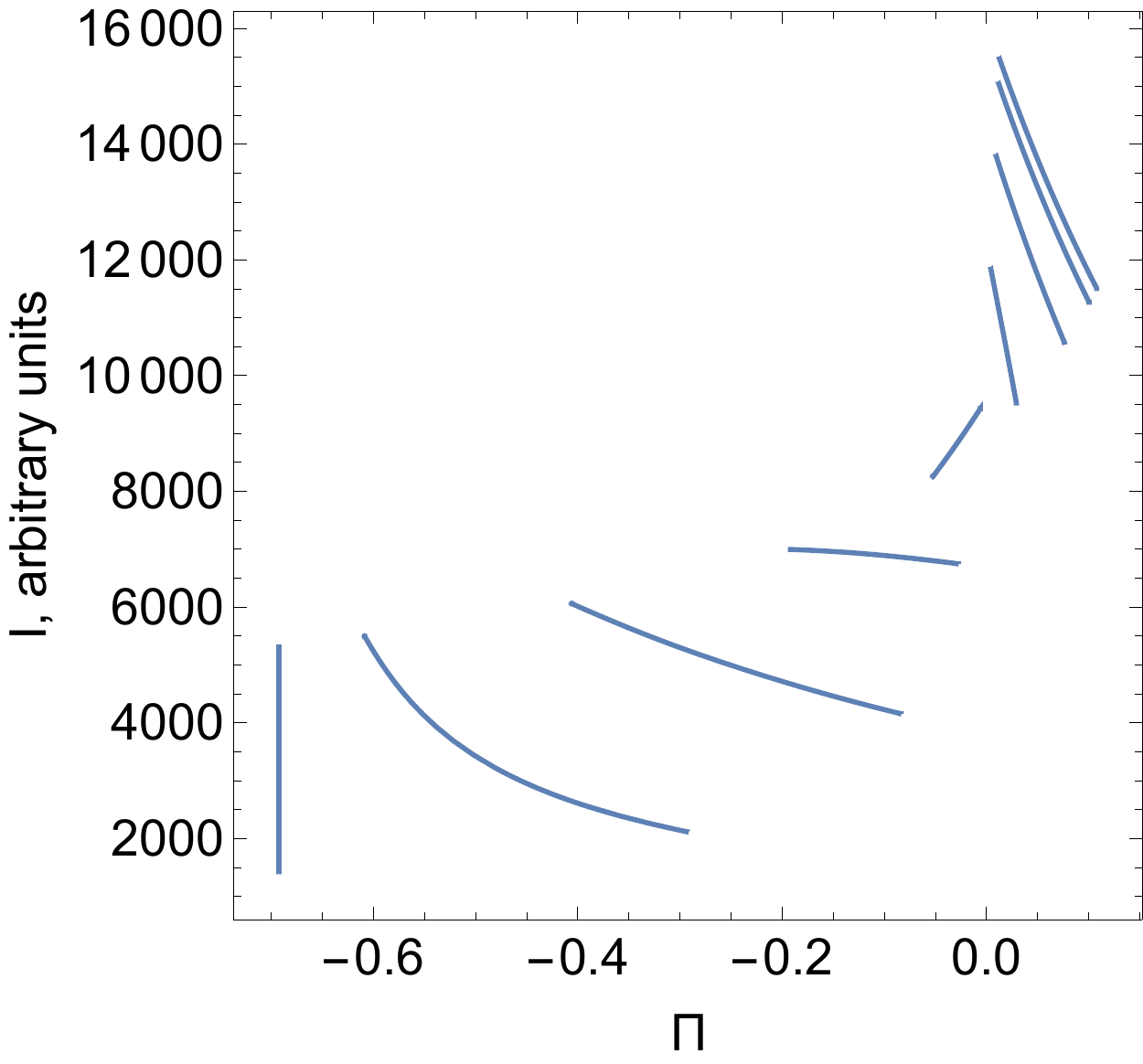}
\caption{Intensity (first row) and polarization (second row) and EVPA (third row) as function of the jet direction for a jet executing a circular motion.  Fourth row: polarization as function of intensity. Jet Lorentz factor is $\gamma=10$, viewing angle $\theta_{ob, min} = 0.02$. The jet makes a circle with opening angle $\theta_j = 0.03$ (left column),  $\theta_j = 0.02$ (center column, so that sometimes the line of sight is directed along the jet) and $\theta_j = 0.01$. The dashed line (second column, third row) indicates that when jet direction passes close to the line of sight the  EVPA actually experiences $\sim \uppi$ swing (it is unobservable for exact alignment). Polarization has minimum and intensity is maximal when the rate of EVPA swing is maximal.} 
\label{fig_rot}
\end{figure*}

Next we consider a jet executing a circular motion. Now, for a given jet \Lf\ and minimal viewing angle $\theta_{ob, 0}$ there is a parameter $\theta_j$ - the size of the circle that the jet's direction makes on the plane of the sky. Results are plotted in Fig. \ref{fig_rot}. A new feature now is that the  EVPA can experience large continuous swings. The total change of EVPA is unlimited for  a jet making several rotations.

%\begin{comment}
%\end{comment}

\subsection{Overall trends and special cases}
The visual inspection of the modeling results reveals following patterns of the EVPA swings:
\begin{itemize}
\item Fastest swings of EVPA typically coincide with intensity and polarization extrema (maxima or minima).
% (e.g. center and right panels in Fig. \ref{Piofphij}); yet for some parameters the polarization fraction can be minimal at the moment of fast EVPA swing and intensity maxima  (e.g.  left panels in Fig. \ref{Piofphij}). See also last row in Fig. \ref{Piofphij}.

\item  EVPA can experience fast $\pm \uppi/2 $ jumps near $\Pi=0$, Fig. \ref{2piover7}. At these points intensity can be maximal or minimal.

\item Swinging jets can produce very fast EVPA swings, up to  $\pm \uppi$; at the maximum rate of the EVPA swings the polarization fraction minima can be maximal or minimal (first  {columns} in Figs. \ref{Piofphij} and \ref{Piofphij1}).

\item For somewhat similar intensity and polarization behaviors, the behavior of the EVPA can be  high variable: smooth evolution (left panel in Fig. \ref{fig_rot}), a sudden jump (center panel in Fig. \ref{fig_rot}) or an oscillations (right panel in Fig. \ref{fig_rot}). 
Importantly, in most cases the fastest swing of EVPA corresponds to a minimum of polarization fraction and an extremum (maximum or minimum) in intensity. 
Yet, for specific case {-- the case of swinging jet passing along the line of sight  (center column, third row, solid line in Fig.~\ref{fig_rot})} -- no EVPA swing is seen (corresponding to a neatly $\uppi$ swing in EVPA, dashed line in the same panel). 

\item Intensity depends on the  EVPA and the polarization fraction in a complicated way; for accelerating jets such  dependence is not necessarily symmetric, Fig. \ref{PiofgammaS}. 

\item {Swinging jets can give EVPA rotations in opposite directions within the same source (Fig. \ref{Piofphij}, change in oscillating angle from $\uppi$/2 back to $-\uppi$/2).}

\item {Circularly moving and swinging jets can produce EVPA rotations with the amplitude $\leq\uppi/2$, including complex changes (e.g. Fig. \ref{fig_rot}), if a jet experience only slight changes in its direction.}

\end{itemize}

We stress that these very complicated,  apparently random, and yet correlated variations of intensity, polarization fraction and EVPA swings come from the assumed highly regular jet motion and very regular jet structure.
Most importantly, we assumed \textit{constant} jet emissivity. 
We expect that variations in the acceleration/emissivity properties of the jets will further complicate the observed properties. 
As well as more complicated trajectories of a beam can give more complicated profiles.

\section{Comparison with observations}
\label{Comparison}

The number of detected EVPA rotations to date {(at least those which we known)} accounts {52} in 20 blazars (see Table \ref{tb1} and Table \ref{tb2}) and are observed to occur on dramatically different time scales - from days to months.
%\citep[][]{2015MNRAS.453.1669B,2015ApJ...809..130C,2016MNRAS.457.2252B,2016AA...590A..10K} 
%Among these EVPA rotations almost half has amplitude of total EVPA change less than 180 degrees, and the other half shows amplitude change higher than 180 degrees, incuding six events exhibiting rotation larger than 360 degrees. 
The amplitude of a majority of these rotations lies in range from 70$^{\circ}$ to 360$^{\circ}$, while larger rotations have also been observed.

The bulk of these rotations possesses the same properties (though more complex behavior is observed):
\begin{itemize}
\item degree of polarization is highly variable \citep[e.g.][]{2012PASJ...64...58S}, while during the EVPA swing $\Pi$ is lower than during the intervals with no rotations \citep[e.g.][]{2016MNRAS.457.2252B};
\item EVPA changes smoothly and exhibits small variations around some mean value during the quiescent state of the source \citep[e.g.][]{2010Natur.463..919A};
\item the fast changes in EVPA seem to occur when polarization fraction passes through minimum \citep[e.g.][]{2016MNRAS.457.2252B}; 
\item total optical flux experiences flares or doesn't change at all.
\end{itemize}

\begin{table}
\caption{Detected to date EVPA rotations in blazars. See also Table~\ref{tb2}.\label{tb1}}
\begin{tabular}{lcl}
\hline
Blazar name & Amplitude of & Reference\\
(B1950) & rotation (deg) & \\
\hline
PKS~0420$-$014 & $-$110 & \citet{2007ApJ...659L.107D}\\
S5~0716$+$71 & $+$180 & \citet{2013ApJ...768...40L}\\
S5~0716$+$71 & $+$180 & \citet{2015ApJ...809..130C}\\
 & $-$180 & $-\prime\prime-$\\
 & $+$300 & \citet{2016Galax...4...43L}\\
 & $+$300 & $-\prime\prime-$\\
OJ~287 & $-$180 & \citet{1988AA...190L...8K}\\
S4~0954$+$65 & $+$330 & \citet{2014AJ....148...42M}\\
 & $-$330 & $-\prime\prime-$\\
W~Comae & $+$110 & \citet{2013EPJWC..6107010B}\\
3C~279 & $\thicksim$90 & \citet{2010Natur.463..919A}\\
 &  & \citet{2016AA...590A..10K}\\
3C~279 & $-$290 & \citet{2008AA...492..389L}\\
%3C~279 & $+$140 & \citet{2014AA...567A..41A}\\
3C~279 & $-$500 & \citet{2016AA...590A..10K}\\
 & $-$100 & $-\prime\prime-$\\
 & $+$350 & $-\prime\prime-$\\
PKS~1510$-$089  & $+$720 & \citet{2010ApJ...710L.126M}\\
PKS~1510$-$089 & $+$400 & \citet{2014AA...569A..46A}\\
PKS~1510$-$089 & $-$250 & \citet{2011PASJ...63..489S}\\
 & $+$500 & $-\prime\prime-$\\
BL~Lac & $+$220 & \citet{2008Natur.452..966M}\\
BL~Lac & $+$210 & \citet{1993ApSS.206...55S}\\
CTA~102 & $-$180 & \citet{2015ApJ...813...51C}\\
3C~454.3 & $+$130 & \citet{2010PASJ...62..645S}\\
3C~454.3 & $-$500 & \citet{2012PASJ...64...58S}\\
 & $+$400 & $-\prime\prime-$\\
\hline
\end{tabular}
\end{table}

\begin{table}
\caption{Detected EVPA rotations in blazars within RoboPol monitoring program.\label{tb2}}
\begin{tabular}{lcl}
\hline
Blazar name & Amplitude of & Reference\\
(B1950) & rotation (deg) & \\
\hline
OC~457 & $-$225 & \citet{2015MNRAS.453.1669B}\\
 & $-$92 & \citet{2016MNRAS.457.2252B}\\
PKS~0256$+$71 & $-$180 & \citet{2015MNRAS.453.1669B}\\
S5~0716$+$71 & $-$208 & $-\prime\prime-$\\
OJ~287 & $-$154 & $-\prime\prime-$\\
GB6~J1037$+$5711 & $-$165 & \citet{2016MNRAS.457.2252B}\\
S4~1044$+$71 & $-$188 & \citet{2015MNRAS.453.1669B}\\
PKS~1510$-$089& $+$243 & \citet{2016MNRAS.457.2252B}\\
 & $-$200 & $-\prime\prime-$\\
PG~1553$+$113 & $+$128 & \citet{2015MNRAS.453.1669B}\\
 & $+$145 & \citet{2016MNRAS.457.2252B}\\
TXS1557$+$565 & $+$222 & \citet{2015MNRAS.453.1669B}\\
S4~1749$+$70 & $-$126 & \citet{2016MNRAS.457.2252B}\\
OT~081 & $-$335 & $-\prime\prime-$\\
S5~1803$+$784 & $-$192 & $-\prime\prime-$\\
3C~371 & $-$347 & \citet{2015MNRAS.453.1669B}\\
 & $+$238 & $-\prime\prime-$\\
 & $-$187 & \citet{2016MNRAS.457.2252B}\\
S4~1926$+$61 & $-$105 & \citet{2015MNRAS.453.1669B}\\
S5~2023$+$760 & $+$107 & \citet{2016MNRAS.457.2252B}\\
BL~Lac & $-$253 & \citet{2015MNRAS.453.1669B}\\
CTA~102 & $-$312 & $-\prime\prime-$\\
 & $-$140 & $-\prime\prime-$\\
RGB~J2243$+$203 & $-$183 & $-\prime\prime-$\\
3C~454.3 & $-$129 & $-\prime\prime-$\\
 & $+$145 & \citet{2016MNRAS.457.2252B}\\
B2~2308$+$34 & $+$74 & \citet{2015MNRAS.453.1669B}\\
\hline
\end{tabular}
\end{table}

%0716+714 and planar motion
%\begin{comment}
\begin{figure}
 \centering
 \includegraphics[width=.99\columnwidth]{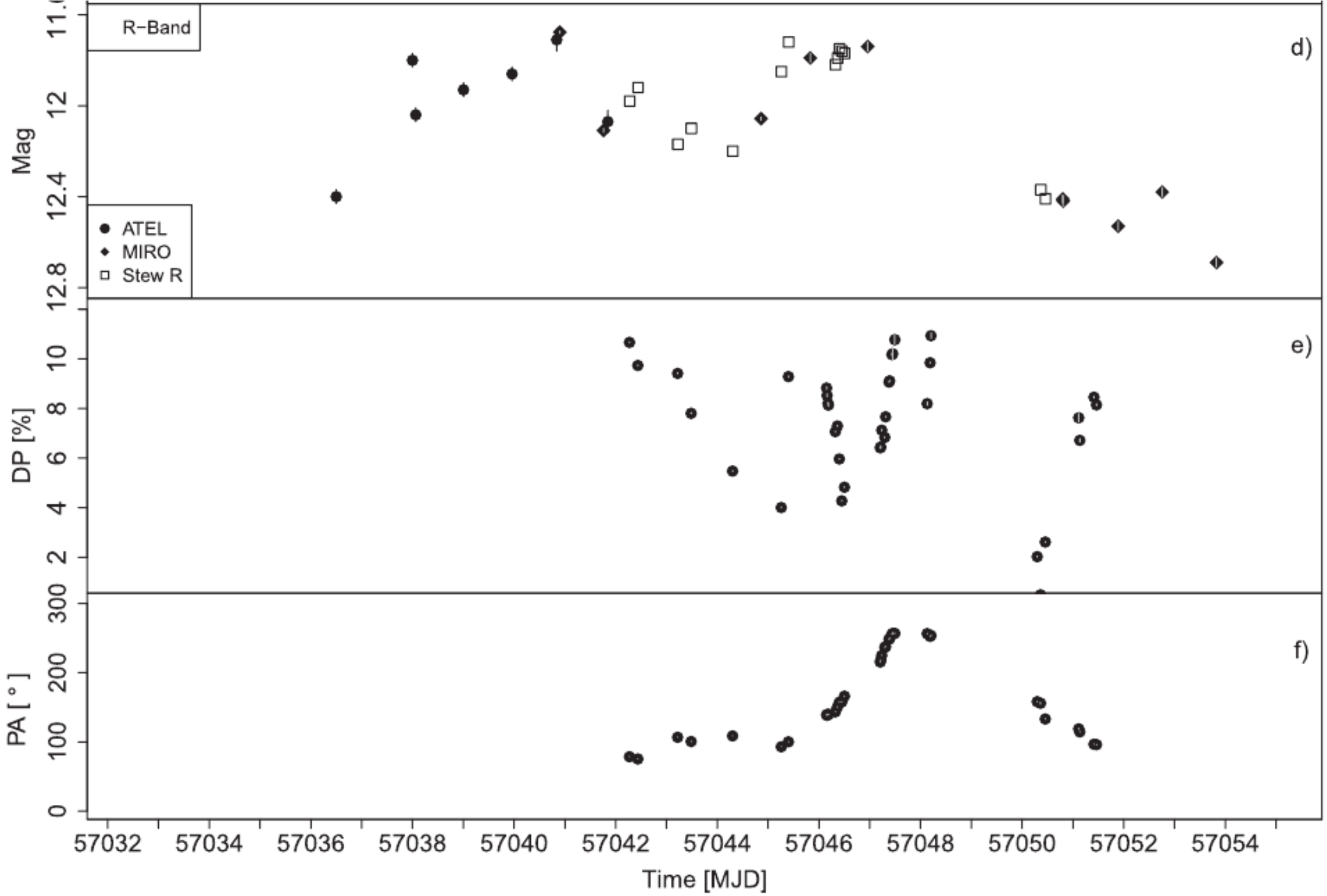}
 \caption{\textit{From top to bottom:} UV/optical magnitudes, degree of polarization and EVPA of S5~0716$+$71 during 2015, January. Figure 1 of \citet{2015ApJ...809..130C}.}
 \label{fig_pa0716c}
\end{figure}
%\end{comment}

\subsection{Fast $\pm180^{\circ}$ rotations}
%{{Fast $\pm180^{\circ}$ rotations. Jet planar motion.}}
Nearly sixty per cents of observed rotational events have maximum amplitude of EVPA change $\lesssim180^{\circ}$.
Since our model explains complex behavior of fast $\pm180^{\circ}$ rotations, we consider here only few EVPA rotational events.
%Among EVPA rotations with the amplitude less than $\thicksim180^{\circ}$ sources follow the pattern, predicted within the model of planar jet motion (\S \ref{Planar}).
One particularly interesting source is S5~0716$+$71, in which four $\pm180^{\circ}$ rotational events have been registered to date \citep{2013ApJ...768...40L,2015ApJ...809..130C,2015MNRAS.453.1669B}.
The complex, seemingly erratic behavior of the degree of polarization in 2015 in the source \citep{2015ApJ...809..130C}, given in Fig.~\ref{fig_pa0716c}, is accompanied by smooth clockwise rotation of the optical EVPA with 180$^{\circ}$, which then changes the direction and rotates back with the same amplitude.
Such behavior, as well as EVPA swings in 2011 \citep{2013ApJ...768...40L} and 2013 \citep{2015MNRAS.453.1669B} may be explained by the proposed model (Fig. \ref{Piofphij}, right row), if S5~0716$+$71 jet experiences regular changes in its orientation.
%Indeed, \citet{2009AA...508.1205B}, \citet{2013AJ....146..120L} and \citet{2014AA...571L...2R} report about temporal changes in the orientation of S5~0716+71 parsec-scale jet, which correlates with the radio and $\gamma$-ray flaring activities (see Fig.~\ref{fig_rani0716}).
Indeed, \citet{2009AA...508.1205B}, \citet{2013AJ....146..120L} and \citet{2014AA...571L...2R} report about S5~0716$+$71 parsec-scale jet variations, showing correlation with the radio and $\gamma$-ray flaring activity \citeauthor{2014AA...571L...2R}
Their analysis supports our idea about the main contribution of a variations in S5~0716$+$71 jet direction to the observed outburst behavior.

%\begin{comment}
\begin{figure}
 \centering
 \includegraphics[width=.82\columnwidth]{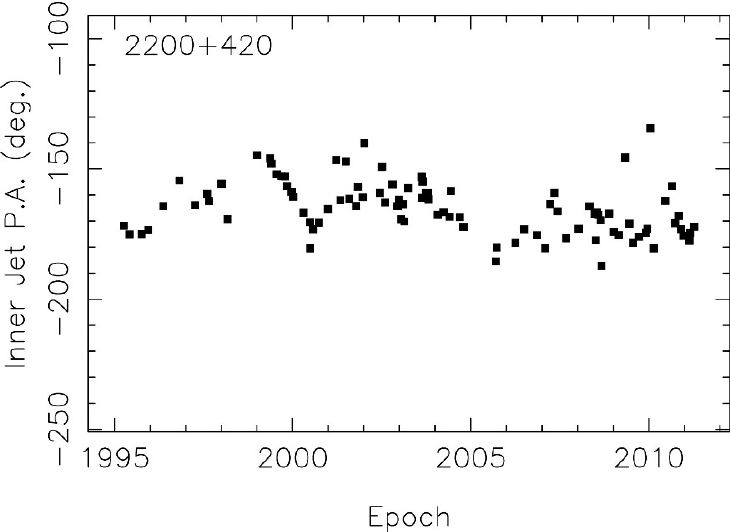}\\
 \caption{Innermost jet position angle vs. time, observed within VLBA MOJAVE program at 15 GHz for BL~Lac. Figure 7 of \citet{2013AJ....146..120L}.}
 \label{fig_pam}
\end{figure}
%\end{comment}

Other interesting source is BL~Lac, which is observed to show fast EVPA swing events ($+210^{\circ}$, \citet{1993ApSS.206...55S}; $+220^{\circ}$, \citet{2008Natur.452..966M}; $-253^{\circ}$, \citet{2015MNRAS.453.1669B}), examples of which are presented in Fig.~\ref{fig_bllac_mars}. 
Observed behavior of these EVPA rotations is very close to the predictions of the proposed model (Fig. \ref{Piofphij}), suggesting structural changes in the jet of BL~Lac.
Indeed, \citet{2003MNRAS.341..405S} and \citet{2013MNRAS.428..280C} show that the BL~Lac jet oscillates on the plane of the sky (see Fig.~\ref{fig_pam}), which supports our assumptions.

%PKS 1510-089
%\begin{figure}
% \centering
% \includegraphics[width=.76\columnwidth]{marscher_fig4.eps}
% \caption{\textit{From top to bottom:} optical flux density,  degree of polarization and EVPA of optical polarization in PKS~1510$-$089 in early 2009. Filled black circles: R band; filled triangles: V band; open squares: $\lambda$ = 500--700 nm. For details see \citet{2010ApJ...710L.126M}.}
% \label{fig_marscher_fig4}
%\end{figure}

\subsection{Large EVPA rotations (jet's circular motion)}
%{Circular motion.}
Other forty per cents of the observed rotational events have maximum amplitude of EVPA change higher than $180^{\circ}$, including six events with the rotation $>360^{\circ}$. 
Deviation of the observed amplitudes of EVPA changes from 180$^{\circ}$ and 360$^{\circ}$ indicates that EVPA can make non-integer number of turns around the line of sight, accompanied by a partial or sequential $\Pi$, EVPA and $I$ patterns, predicted by the model.
%\bf{Circular motion. EVPA rotations $>$360$^{\circ}$.}
Jets of, for example, CTA~102 \citep{2015MNRAS.453.1669B,2015ApJ...813...51C}, TXS~1557$+$565 \citep{2015MNRAS.453.1669B}, 3C~371 \citep{2015MNRAS.453.1669B}, 3C~279 \citep{2016AA...590A..10K} and others, might be considered to exhibit circular motion.

%\begin{comment}
\begin{figure}
 \centering
 \includegraphics[width=.99\columnwidth]{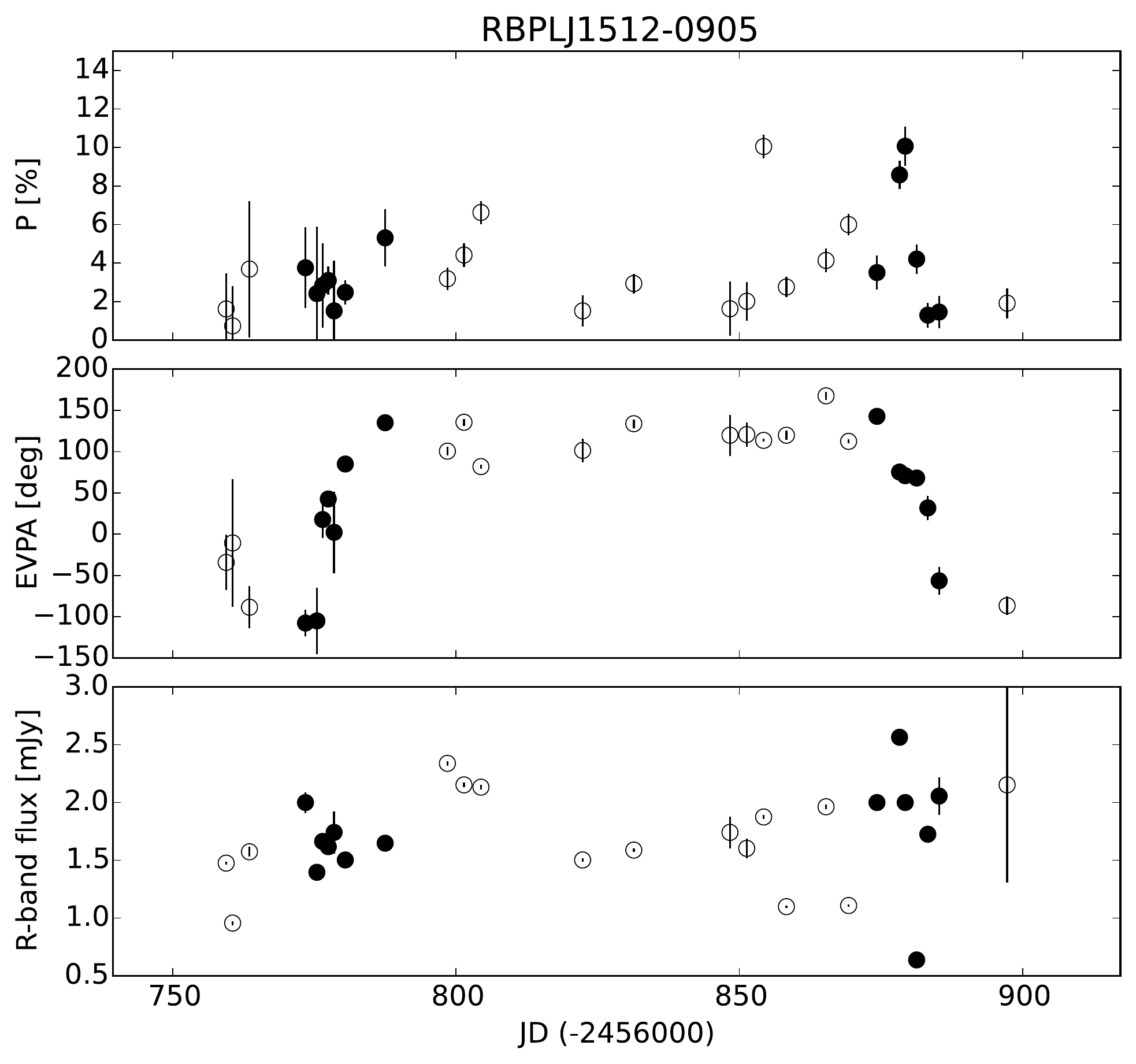}
 \caption{\textit{From top to bottom:} evolution of polarization degree, EVPA and R-band magnitude for PKS~1510$-$089 with a detected EVPA rotation in the RoboPol monitoring program. Periods of rotations are marked by filled black points. Figure 3 of \citet{2016MNRAS.457.2252B}.}
 \label{fig_pks15_blin}
\end{figure}
%\end{comment}

The largest EVPA swing to date is observed in PKS~1510$-$089 \citep{2010ApJ...710L.126M} and amounts 720$^{\circ}$. %(Fig.~\ref{fig_marscher_fig4}).
%, with the indication of a single emission feature is responsible for the rotation (Fig.~\ref{fig_marscher_fig4}).
Assuming circular motion of the blazar jet, this overall rotation might be represented by a sequence of 180$^{\circ}$--360$^{\circ}$ EVPA turns.
Other five detected rotational events in this blazar \citep{2011PASJ...63..489S,2014AA...569A..46A,2015MNRAS.453.1669B,2016MNRAS.457.2252B} have amplitudes from $200^{\circ}$  to 500$^{\circ}$ and occur in different directions (Fig.~\ref{fig_pks15_blin} shows an example of blazar activity in 2014). 
%The axample of blazar activity in 2014 is given in Fig.~\ref{fig_pks15_blin}.
These continuous EVPA variations imply that the PKS~1510$-$089 jet executes irregular, circular motion.
Different emission features, rotating in opposite directions relative to the line of sight, or close proximity of the jet viewing angle to the LoS may be responsible for the complex blazar flaring behavior in time.

%3C279
%\begin{comment}
\begin{figure*}
 \centering
 \includegraphics[width=1.40\columnwidth]{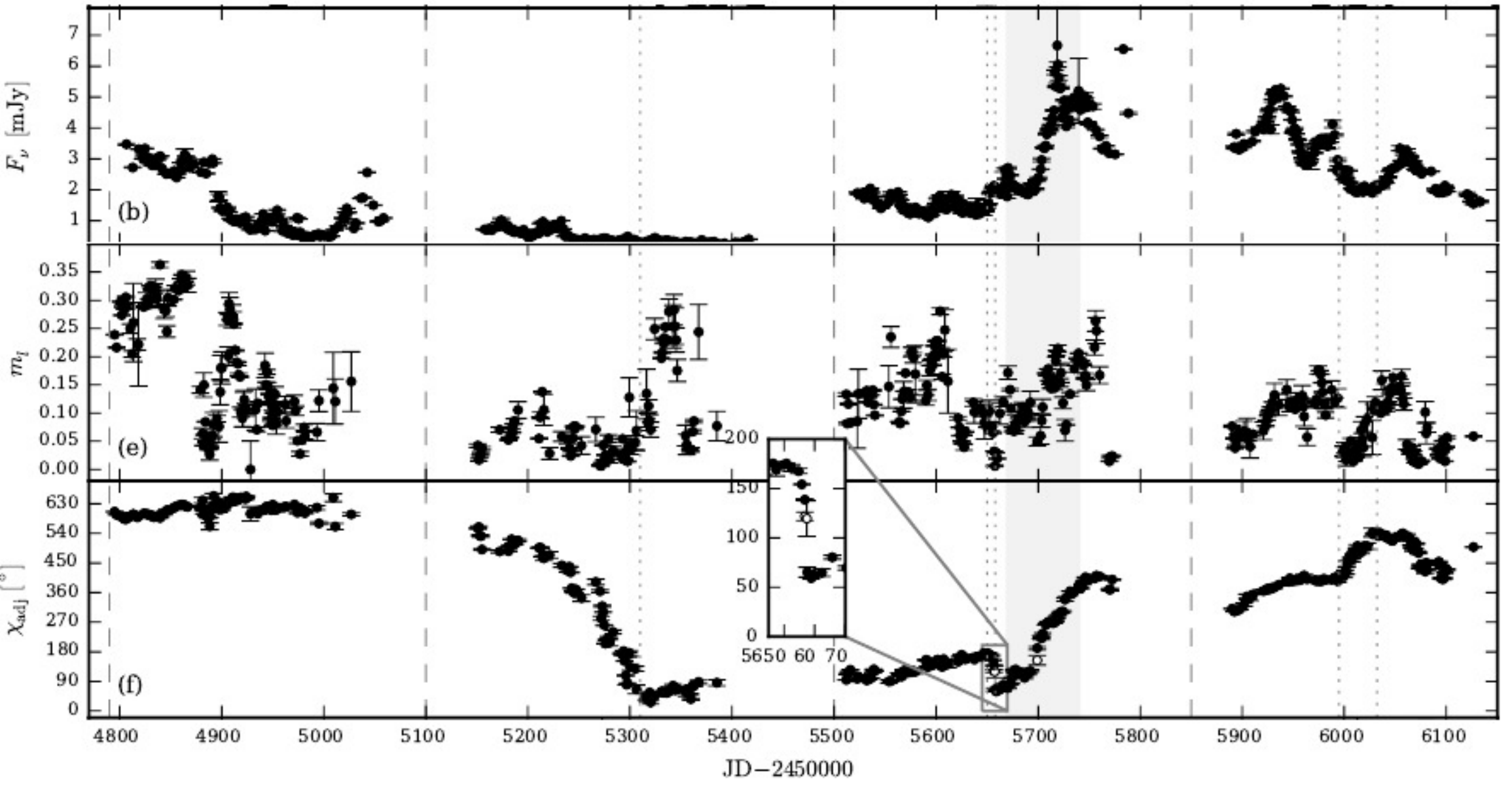}
 \caption{Optical photometry and polarimetry of 3C~279. (b) Combined R-band light curve. (e) Combined, de-biased, and averaged polarization fraction. (f) Combined, averaged, and adjusted EVPA; open symbols are added from the non-averaged EVPA curve. The grey area highlights the period of $\gamma$-ray flaring activity coinciding with a rotation of the optical polarization angle. Figure 1 of \citet{2016AA...590A..10K}.}
 \label{fig_pa_3c279}
\end{figure*}
%\end{comment}

%3C454.3 No figures
Blazar 3C~454.3 also experiences large EVPA rotational events (see Table~\ref{tb1} and Table~\ref{tb2}). 
Its jet shows complex VLBI structure \citep[e.g.][]{2005AJ....130.1418J} with the indication of a very small angle of jet direction to the line of sight.
This may explain complex overall behavior of its properties during both active and quiescent states. 
Considering the 3C~454.3 jet executing a circular motion, which in a sky projection turns into planar motion in some cases, our model may explain its observed overall properties: fast EVPA swings with the amplitude less than 180$^{\circ}$ \citep{2010PASJ...62..645S,2015MNRAS.453.1669B,2016MNRAS.457.2252B} as well as large and continuous EVPA rotations \citep[$-$500$^{\circ}$ and $+$400$^{\circ}$,][] {2010PASJ...62..645S,2012PASJ...64...58S}.

\begin{figure}
 \centering
  \includegraphics[width=.89\columnwidth]{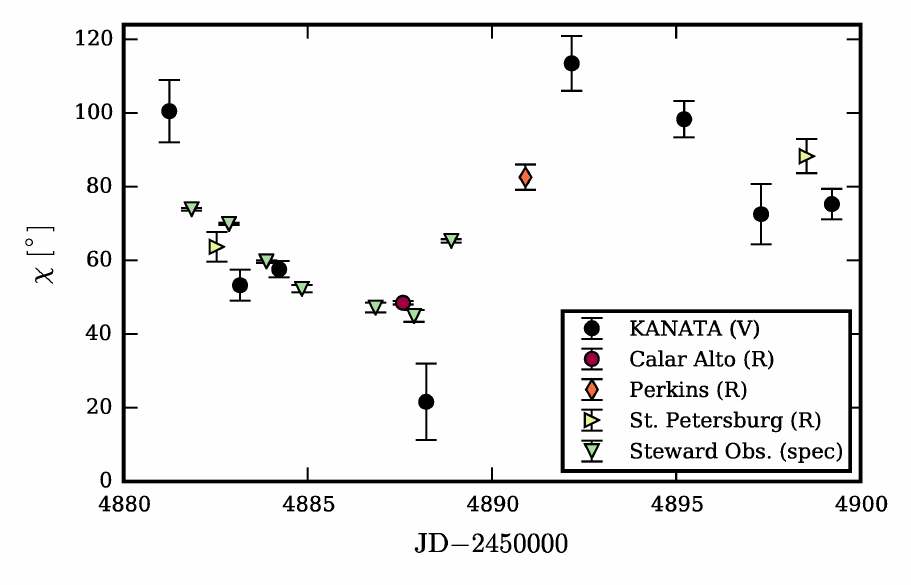}\\
  \includegraphics[width=.8\columnwidth]{PARotatingThetaob-hundreths.pdf}
 \caption{\textit{Top panel}: The EVPA rotation in the jet of 3C~279 in 2009 (Fig. 3 of \citet{2016AA...590A..10K}). \textit{Bottom panel}: EVPA (third row right column of Fig. \ref{fig_rot}) as function of the jet direction for a jet executing a circular motion. There is apparent qualitative agreement - no fits/parameter fine tuning has been done.}
 \label{fig_pa_3c279ab}
\end{figure}

Another source exhibiting a number of different EVPA rotations, is 3C~279. 
%An example of a 3-year optical photometry and polarimetry of 3C~279 is given in Fig.~\ref{fig_pa_3c279}.
The largest rotation in its jet amounts 500$^{\circ}$ \citep[counterclockwise rotation, ][]{2016AA...590A..10K} and occurs during the quiescent state of the source, shown in Fig.~\ref{fig_pa_3c279}. 
\citet{2016AA...590A..10K} conclude that the stochastic process is responsible for this rotation.
We argue, that the model of a circular motion of a jet can reproduce observed properties of this EVPA rotational event (Fig. \ref{fig_rot}), resulted in relatively constant optical flux, zero degree of polarization and continuos change of EVPA.
%Prolonged monitoring of 3C~279 \citep{2016AA...590A..10K} shows (see Fig.~\ref{fig_pa_3c279}), that besides the possible circular motion, its jet experiences other variations, accompanied by  fast  erratic changes of  EVPA. 
The small, short-term fluctions overlaid on this 500$^{\circ}$ rotation may arise from the turbulent component of magnetic field, superimposed with a large scale ordered magnetic field.
Or being due to small-amplitude random short-term ``jitter'' of the jet's direction, superimposed on a continuos longer term evolution. 
Properties during the other EVPA rotations in this source \citep{2008AA...492..389L,2014AA...567A..41A,2016AA...590A..10K} also can be interpreted within proposed model, including the complex $\thicksim90^{\circ}$ rotation in 2009 \citep{2010Natur.463..919A,2016AA...590A..10K}, represented in Fig.~\ref{fig_pa_3c279ab}.
%Position angle swings of opposite direction may result from the different emission features, rotating in opposite directions relative to the line of sight.

%\begin{comment}
%\end{comment}

%{Cases of complex EVPA rotations.}
\subsection{Complex EVPA rotations}
\begin{figure*}
 \centering
 \includegraphics[scale=4.54]{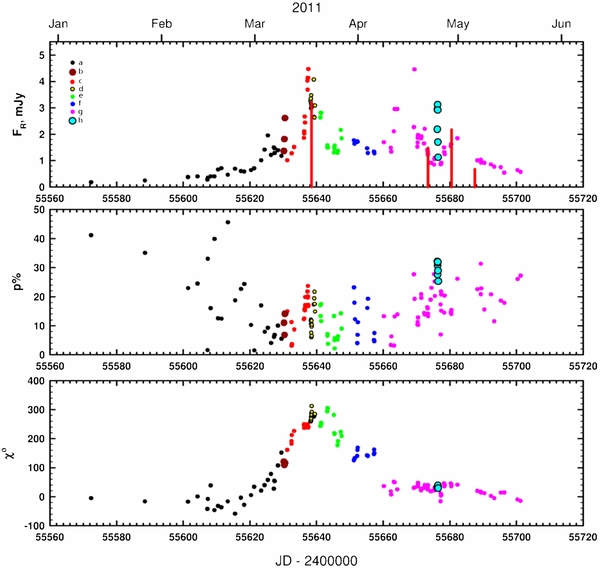}
 \caption{Optical flux density (corrected for Galactic extinction), fractional polarization, and position angle of polarization in R band vs. time in 2011 January--May in  S5~0954$+$65, Figure 2 of \citet{2014AJ....148...42M}. }
 \label{fig_pa_s41156m}
\end{figure*}

In some cases the EVPA shows complex behavior, as discussed above 3C~279. 
%S5~0954+65
The other example is the flare in S5~0954$+$65 in 2011 \citep{2014AJ....148...42M}, Fig. \ref{fig_pa_s41156m}.
EVPA steadily increases from 0$^{\circ}$ to 330$^{\circ}$ and then recovers back to 0$^{\circ}$, accompanied by strong variations in optical flux density and fractional polarization.
\citet{2014AJ....148...42M} explains such activity by a superposition of two radiative components, with  only one component is responsible for EVPA rotation.
Applying this explanation to the predictions of our model, one may expect, that the jet of S5~0954+65 experiences oscillating circular motion.

%\begin{comment}
%\end{comment}

\section{Discussion}
In this work we discuss a model  of blazar activity - a jet  carrying helical \Bf\ with a regularly changing direction (and possibly changing bulk \Lf) and constant emissivity. 
We demonstrate that this highly deterministic model can produce highly variable polarization,  EVPAs, and intensity profiles. 
At the same time the model reproduces smoothly varying EVPA changes.
Thus,  though for any given configuration the intensity, polarization and the EVPA are deterministic and thus  their behavior is highly correlated, the non-monotonic variations of these values as functions of the jet direction and \Lf\ produce highly variable overall behavior.
We find that for smooth variation of EVPA (i) $\Pi$ can be highly variable; (ii) $\Pi$ is close to zero at the moment of fastest EVPA swing; (iii) the intensity is usually maximal at points of fastest EVPA swing, but can have a minimum; (iv) for some special pitch angles there are large fluctuations of EVPA, but this always occurs at small $\Pi$. 
%\color{red}{Delete? These results are highly encouraging. This simple model explains the large variety of EVPA swings...} For a simple linear trajectory of a beam the model can reproduce both overall correlations, like small polarization during maximal rate of EVPA swings, as well as often-observed exceptions.

Importantly, these features are obtained for the assumed \textit{constant} intrinsic jet emissivity. Variations in the acceleration/emissivity properties of the jets, more complicated or irregular beam trajectories, as well as existence of multiple emission components within one beam will complicate observed profiles. 
In addition to the possible presence of a turbulent magnetic fields, these features may produce small ($<90^{\circ}$), seemingly random EVPA variations \citep[e.g.][]{2013ApJ...768...40L}.  
Such variations are often observed during the quiescent state of the source and are not considered in this paper.

Previously, a number of models have been proposed to explain blazar variability, such as propagation of the emission region along a helical \citep{2010Natur.463..919A}, spiral \citep{2010ApJ...710L.126M} or bent trajectories \citep{2014AA...566A..26M}, relativistic shocks propagating in a jet \citep{1985ApJ...298..114M}, interaction of the relativistically moving plasma with a standing shocks \citep{2014ApJ...780...87M}, magnetic reconnection \citep{2003NewAR..47..513L,2009MNRAS.395L..29G} and passage of a disturbance through the emission region of a jet \citep{2014ApJ...789...66Z}.
Unlike above mentioned works, the main idea of the proposed model is to interpret observed polarization and kinematic signatures of the blazar jets and not to consider physical background of the jet emission properties and mechanisms, like particle energy distribution, acceleration and $\gamma$-ray production.

The fact that EVPA rotations have been detected only in $\gamma$-ray-loud objects suggests the better alignment of the jet-emitting regions with the line of sight during these events or large apparent speeds in these objects \citep[e.g.][]{2015ApJ...810L...9L}, resulting in increase of Doppler boosting factor. 
One may expect that within the proposed model  the strongest $\gamma$-ray flares are  accompanied by the fastest EVPA rotations.

We envision that changing jet direction is due to the changing launching direction, so that the motion of each element is (nearly) ballistic.  
The variability is then created not at the central \BH\, but within the flow itself. 
For example, we may be seeing mini-jets
\citep{2006MNRAS.369L...5L,2009MNRAS.395L..29G,2008MNRAS.386L..28G,2009MNRAS.395..472K} 
that move relativistically {within} the overall jetted outflow.  
Relativistic internal sub-jets can result from reconnection occurring in highly magnetized plasma \citep[]{2003ApJ...589..893L,2012SSRv..173..521H} due to the development of current-driven instabilities \citep[e.g.][]{2016MNRAS.456.1739B}.
The corresponding time scales in the jet frame are the dynamical (\Alfven) time of the inner part of the jet. In the observer frame the time scale will be modified by the relativistic effects,  and can be considerably shorter, by a factor $1/\gamma^2$.

We would like to thank Tobias Beuchert, Sebastian  Kiehlmann, Matthew Lister, Alexander Pushkarev, Hao-Cheng Zhang  and the participants of the Jets2016 meeting\footnote{\url{http://jets2016.iaa.es}}. ML  was supported by NSF grants AST-1306672 and AST-1516958.

\bibliographystyle{mnras}
% \bibliography{/Users/maxim/Home/Research/BibTex}  \end{document}
\bibliography{liter2}

\bsp
\label{lastpage}
\end{document}